\documentclass{aa}
\usepackage{graphicx}
\usepackage{txfonts}
\usepackage{multirow}
\usepackage{natbib,twoopt}
\bibpunct{(}{)}{;}{a}{}{,} % to follow the A&A style
\usepackage{adjustbox} 
\usepackage{graphicx}
\usepackage{lscape}
\usepackage{wrapfig}
\usepackage{subfigure}
\DeclareGraphicsExtensions{.ps}
\usepackage{booktabs}
\usepackage{adjustbox} 

\makeatletter
\newcommandtwoopt{\citeads}[3][][]{\href{http://adsabs.harvard.edu/abs/#3}%
{\def\hyper@linkstart##1##2{}%
\let\hyper@linkend\@empty\citealp[#1][#2]{#3}}}
\newcommandtwoopt{\citepads}[3][][]{\href{http://adsabs.harvard.edu/abs/#3}%
{\def\hyper@linkstart##1##2{}%
\let\hyper@linkend\@empty\citep[#1][#2]{#3}}}
\newcommandtwoopt{\citetads}[3][][]{\href{http://adsabs.harvard.edu/abs/#3}%
{\def\hyper@linkstart##1##2{}%
\let\hyper@linkend\@empty\citet[#1][#2]{#3}}}
\newcommandtwoopt{\citeyearads}[3][][]%
{\href{http://adsabs.harvard.edu/abs/#3}
{\def\hyper@linkstart##1##2{}%
\let\hyper@linkend\@empty\citeyear[#1][#2]{#3}}}
\makeatother

\begin{document}
   \title{High-resolution X-ray spectroscopy of Cen X$-$3 with XMM-Newton}

   \author{J. J. Rodes-Roca\inst{1,2} \and
          J. M. Torrej\'on\inst{1,2} \and G. Sanjurjo-Ferr\'{\i}n\inst{2} \and J. Planelles Villalva\inst{2}
          }

   \offprints{J. J. Rodes-Roca}

   \institute{Department of Physics, Systems Engineering and Signal Theory, University of Alicante,
              03080 Alicante, Spain\\
              \email{jjrodes@ua.es}\\
         \and
              University Institute of Physics Applied to Sciences and Technologies,
              University of Alicante, 03080 Alicante, Spain
                          }

   \date{Received     ; accepted       }

  \abstract
   {}
   {The spectral analysis of two \emph{XMM-Newton} observations of the high-mass X-ray binary system Cen X$-$3 is presented. In particular, it is focused on the eclipse and out-of-eclipse spectra in order to compare the properties of the environment around the compact object.}
   {The high-resolution spectra obtained from the reflection grating spectrometer on board \emph{XMM-Newton} was analysed focusing on studying eclipse and out-of-eclipse spectra separately. Several continuum models were explored in \emph{SPEX} for which we studied the properties of emitting and absorbing matter depending on the emission and absorption lines identified in the spectra.}
   {It was found that the X-ray continuum is heavily absorbed by a neutral gas and photoionised matter. Emission lines from Si \textsc{v}, Mg \textsc{xii}, Mg \textsc{xi}, and Ne \textsc{x} were detected in the eclipse spectrum. In particular, H-like lines of Mg and Ne with a significance $\gtrsim5\, \sigma$ in the eclipse spectrum and $\sim 3\, \sigma$ in the out-of-eclipse spectrum. But in the out-of-eclipse spectrum any absorption lines, if any, were detected with a significance less than $2\, \sigma$. \emph{RGS} light curve showed dips in the out-of-eclipse spectrum which are not due to an increase in the column absorption but may be produced by instabilities in the accretion stream. On the other hand, the level of counts above 20 \AA\, was compatible with the X-ray background. A simple local continuum model was used to describe the He-like triplet of Ne and the derived values of $\mathcal{R}$ and $\mathcal{G}$ ratio parameters pointed out that the UV photospheric field should be important at the line production site and an electron density greater than $10^{12}$ cm$^{-3}$. As a consequence, a hybrid plasma may be present in the binary system.}
   {}

   
   \authorrunning{J. J. Rodes-Roca et al.}
   \titlerunning{\emph{RGS} spectra of Cen X$-$3}
   \maketitle
%
%________________________________________________________________

\section{Introduction}

Cen X$-$3 is a high-mass X-ray binary (HMXB) formed by a neutron star orbiting the O6-8 III counterpart V779 Cen \citep{1979ApJ...229.1079H}. The eclipsing nature of the system was discovered by \citet{1972ApJ...172L..79S}. Eclipse duration is around 22\% of the orbital period. The high X-ray luminosity \citep{1983ApJ...270..711W} compared with other HMXBs, such as Vela X$-$1 or 4U 1538$-$52, is explained via a combination of stellar wind accretion plus an accretion disc fed by Roche-lobe overflow \citep{1986A&A...154...77T}. The accretion produces copious X-ray emission that photoionises the circumstellar wind producing emission lines due to recombination in highly ionised plasma \citep{1993ApJ...403..322D}. 
 During eclipse, the direct emission from the neutron star is blocked by the donor star and the scattered emission will be enhanced assuming that the ionised plasma is extended at larger distances from the compact object. Exploring the eclipse spectrum using high-resolution X-ray spectroscopy provides a powerful tool to study the stellar wind characteristics. \citet{1996PASJ...48..425E} 
obtained an observation, covering the entire eclipse taken with the \emph{ASCA} satellite. They reported the presence of emission lines from hydrogenic ions of Ne, Mg, Si, S and Fe and studied the variation of the features at different orbital phases. On the other hand, \citet{2003ApJ...582..959W} also studied an eclipse observation taken with the \emph{Chandra} observatory. They detected hydrogen-like emission lines such as Si \textsc{xiv}, Mg \textsc{xii} or S \textsc{xvi}, and helium-like triplets, such as Mg \textsc{xi}, S \textsc{xv}, Si \textsc{xiii} or Fe \textsc{xxv}, although only the two last ones were clearly resolved (see also \citet{2005ApJ...634L.161I}).

In this work we report on the analysis of both emission and absorption lines in the spectra of Cen X$-$3 obtained with the \emph{XMM-Newton} Reflection Grating Spectrometer (\emph{RGS},  \citeauthor{2001A&A...365L...7D} 2001). 
A detailed analysis of the \emph{EPIC} spectra was presented in \citet[][Paper I]{2021MNRAS.501.5892S}. Prominent emission lines from highly ionised Fe (Fe \textsc{xxv} He-like and Fe \textsc{xxvi} H-like Ly $\alpha$) were present along with Fe K$\alpha$ fluorescence from near neutral Fe. The model used to describe the broadband spectrum was semi phenomenological and composed by a blackbody plus a powerlaw while the emission lines were modelled using Gaussians. This allowed to track the line changes throughout the orbital phase. Here, we are interested in the high resolution spectra provided by the \emph{RGS} instrument with the aim of testing self consistent theoretical plasma models using the rich emission line spectrum below 2 keV.  

The paper is organized as follows: in Sect.~\ref{data} the observations and data reduction are presented. The models and the results of the analysis are presented in Sect.~\ref{analyse}. The discussion and the comparison with previous work are given in Sect.~\ref{discussion} and conclusions are reported in Sect.~\ref{conclusion}.

\section{Observations and data reduction}
\label{data}

Two archival observations of Cen X$-$3 taken with \emph{XMM-Newton} (see Table~\ref{tab:data}) were used. The data were reduced with the \emph{XMM-Newton} Science Analysis System (SAS) version 16.0.0.

\begin{table*}
\caption{\emph{XMM\_Newton RGS} observations used in this paper.}
\label{tab:data}
\centering
\begin{tabular}{lllccr}
\midrule\midrule
ID & Instrument & Mode & Date$_\mathrm{start}$ & Date$_\mathrm{end}$ & Duration (s) \\
\midrule
0111010101 & RGS1 & Spec HER + SES & 2001-01-27T08:11:06 & 2001-01-28T03:45:22 & 65\,922 \\
0111010101 & RGS2 & Spec HER + SES & 2001-01-27T08:11:06 & 2001-01-28T03:45:30 & 63\,569 \\
0400550201 & RGS1 & Spec HER + SES & 2006-06-12T22:07:40 & 2006-06-13T20:36:13 & 80\,370 \\
0400550201 & RGS2 & Spec HER + SES & 2006-06-12T22:07:43 & 2006-06-13T20:36:17 & 80\,370 \\
\hline
\end{tabular}
\end{table*}

\begin{figure*}[h!tb]
  \centering
  \includegraphics[angle=-90,width=\columnwidth]{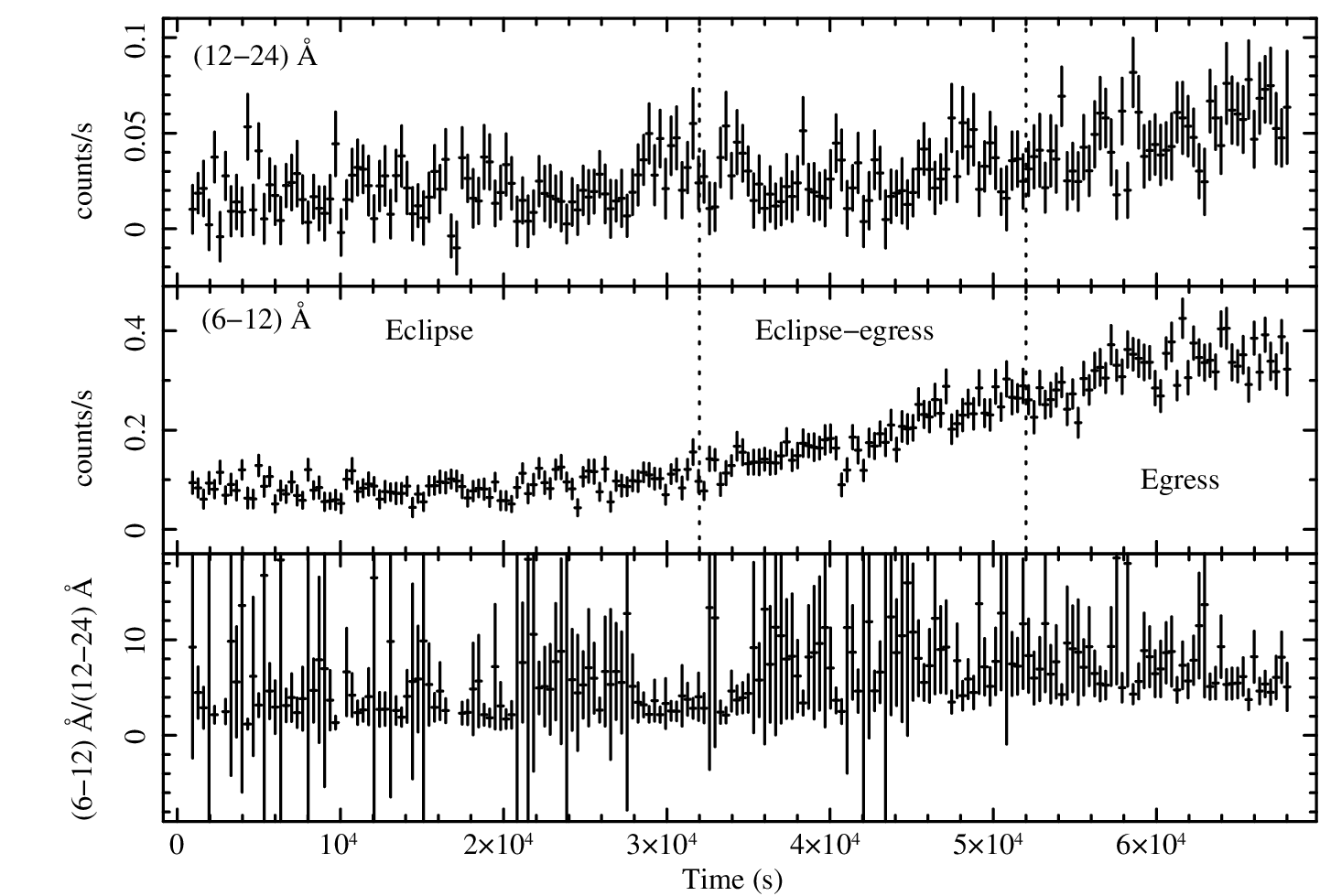}
  \includegraphics[angle=-90,width=\columnwidth]{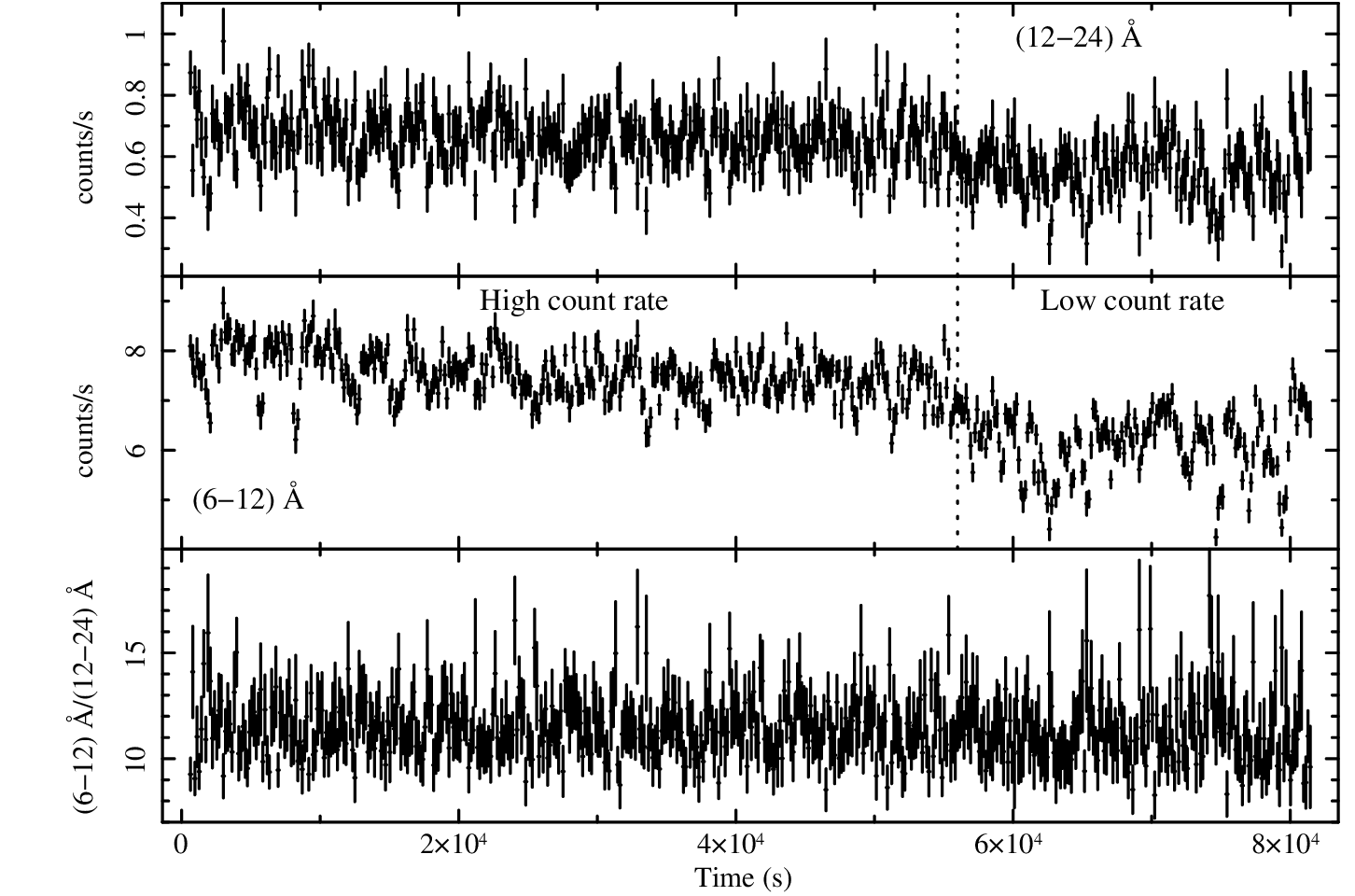}
  \caption{Light curves and hardness ratio of the first (0111010101) low state observation (\emph{left}) and the second (0400550201) high state (\emph{right}) observation with \emph{RGS}. Intervals with high background have already been taken out.
  }
  \label{fig:lcurve}
\end{figure*}

The \emph{RGS} data were processed with the SAS task \emph{rgsproc}. Light curves for the background in CCD9 were obtained following the XMM-SAS procedure\footnote{https://www.cosmos.esa.int/web/xmm-newton/sas-thread-rgs} in order to remove soft proton flares and spurious events. Good time intervals (GTI) were created for filtering periods of high background. Subsequently, we reprocessed the data again with \emph{rgsproc} by filtering them with the GTI for background screening. Response matrices and spectra for the two observations were also extracted. The final net exposure times are reported in Table~\ref{tab:data}. After verifying that the \emph{RGS} 1 and \emph{RGS} 2 were compatible with each other in the whole wavelength range, fluxed spectra for both \emph{RGS} instruments were obtained and stacked them with the SPEX\footnote{htpp://www.sron.nl/spex} task \emph{RGS\_fluxcombine} according to the procedure described by \citet{2011A&A...534A..37K}. Useful data was obtained for the interval (7--38) \AA. Both first order spectra were flux combined into an \emph{RGS} spectrum.
We created the light curves exposure-corrected and background-subtracted using the SAS task \emph{rgslccorr}. Fig.~\ref{fig:lcurve} shows the count rate of the combined \emph{RGS} data in different wavelength bands and the corresponding hardness ratio as a function of time. Observation 0111010101, taken at the source low state (\emph{left}) shows clearly the eclipse egress, more noticeable at the lower energies. The ratio (6--12) \AA/(12--24) \AA\, does not show any noticeable change, although the uncertainties in the eclipse light curve are high. The second observation, 0400550201, taken during a high state and totally out of eclipse (\emph{right}), is more constant but shows a flux decrease at the end. In order to shed light on the changes in the reprocessing environment of the binary system, we performed time resolved spectroscopy during the eclipse ([0--32] ks), eclipse-egress ([32--52] ks) and egress for the observation ID 0111010101. On the other hand, we extracted the \emph{high count rate} ([0--56] ks) and \emph{low count rate} ([56--80] ks)spectra, respectively, for the observation ID 0400550201. We present our results in Sect.~\ref{orbitspec}.

\section{Spectral analysis}
\label{analyse}

The spectral analysis was carried out using the
\emph{SPEX} software version 3.08.01 \citep{1996uxsa.conf..411K}. The interstellar abundances were scaled to the recommended proto-Solar or solar system values of \citet{2009M&PSA..72.5154L}. As the spectra was created with \emph{RGS\_fluxcombine}, the natural Poissonian fluctuations on the spectrum as a function of detector pixel are distributed over the wavelength bin coinciding most with the detector pixel and its neighbours. Due to this tempering of the Poissonian fluctuations, $\chi^2$ values will be lower than C-statistics for the best spectral fit. Therefore, we used the $\chi^2$-statistic and adopted 1 $\sigma$ errors when fit the combined \emph{RGS} spectrum. We assumed the classical distance to the source as $\sim$8 kpc \citep{1974ApJ...192L.135K}, although \citet{2009ApJ...691.1744T} derived a marginally inconsistent distance of 5.7$\pm$1.5 kpc studying the X-ray halo of this object. This assumption is rather consistent with the upper limit of the distance to this source derived from Gaia\footnote{htpp://www.cosmos.esa.int/gaia} by using the parallax measure, the source's G-band magnitude and BP$-$RP colour $6.8^{+0.6}_{-0.5}$ kpc, called by \citet{2021AJ....161..147B} ``photo-geometric'' distance. Spectra were also rebinned by a factor of two, i.e. about 1/3 FWHM (the first order \emph{RGS} spectra provide a resolution of (0.06-0.07) \AA). This gives the optimal binning for the \emph{RGS} and a bin size of about 0.02 \AA.

From the \emph{XMM-Newton/PN} light curve \citep{2012BASI...40..503N,2021MNRAS.501.5892S}, the orbital phase range $\phi = (+0.00-0.37)$ was derived for the first observation. Using the ephemeris of Cen X$-$3 obtained by \citet{2015A&A...577A.130F} and the previous mid-eclipse time, we estimated the orbital phase range $\phi = (0.35-0.80)$ for the second one. Therefore, the spectrum covering the eclipse could be compared with the spectrum out-of-eclipse. First, it was noticed that the spectra above 20 \AA\, show the same level of counts for the source and for the X-ray background. Therefore, the continuum emission was defined using a shorter wavelength range (7--20) \AA\, than that provided by the instruments.

Cen X-3 is a complex system with different X-ray sources, such as the accretion disk, the hot spots on the NS surface or the scattered photons in the strong stellar wind. Consequently, several continuum models were tested for a reasonable description of the averaged spectra modified by neutral gas absorption. Then, the observation ID 0111010101 is divided into three different intervals eclipse, eclipse-egress and egress described in Fig.~\ref{fig:lcurve} (left). Fit results are discussed in Section~\ref{orbphase1}. On the other hand, the observation ID 0400550201 is divided into two intervals high count rate and low count rate described in Fig.~\ref{fig:lcurve} (right). Fit results are discussed in Section~\ref{orbphase2}. It is also performed a separate spectral analysis of each dip detected as well as all of the out-of-dip regions between them marked in Fig.~\ref{fig:dips-nodips}.

\begin{table}[h!tb]
\caption{Phase-averaged spectral parameters for the 0111010101 observation.
}
\label{tab:ecl-pion}
\centering
\begin{tabular}{llc}
\midrule
\midrule
Component & Parameter & Averaged \\
\midrule
\multirow{2}{*}{\emph{Hot$_\mathrm{intrinsic}$}} & $N_H$ (10$^{26}$ m$^{-2}$) & $1^{+5}_{-1}$ \\
 & $k\,T$ (eV) & $127^{+5}_{-6}$ \\ \midrule
\multirow{4}{*}{\emph{PL}} & Norm (10$^{44}$ ph\,s$^{-1}$\,keV$^{-1}$) & $0.77^{+0.01}_{-0.03}$ \\
 & $\Gamma$ & $1.660^{+0.000}_{-0.002}$ \\
 & Flux (10$^{-15}$ W\,m$^{-2}$) & $1.84^{+0.02}_{-0.07}$  \\
 & Unabs. Flux (10$^{-14}$ W\,m$^{-2}$) & $1.31^{+0.02}_{-0.05}$ \\ \midrule
\multirow{2}{*}{\emph{etau$_\mathrm{low}$}} & $\mathrm{tau}_0$ (at 1 keV, frozen) & $1.3605\times 10^{-2}$ \\
 & $a$ (index, frozen) & $-1.00$ \\ \midrule
\multirow{2}{*}{\emph{etau$_\mathrm{high}$}} & $\mathrm{tau}_0$ (at 1 keV, frozen) & $3.3333\times 10^{-3}$ \\
 & $a$ (index, frozen) & $1.00$ \\ \midrule
\multirow{7}{*}{\emph{pion}} & $N_H$ (10$^{26}$ m$^{-2}$) & $3.6^{+0.3}_{-0.7}$ \\
 & $\log \xi$ ($10^{-9}$ W\,m) & $1.576^{+0.011}_{-0.010}$ \\
 & Abundance Ne & $1.27^{+0.12}_{-0.14}$ \\
 & Abundance Mg & $1.9^{+0.3}_{-0.0}$ \\
 & Abundance Fe & $0.5^{+0.4}_{-0.1}$ \\
 & Flux (10$^{-16}$ W\,m$^{-2}$) & $5.8^{+0.5}_{-1.1}$  \\
 & Unabs. Flux (10$^{-15}$ W\,m$^{-2}$) & $7.9^{+0.7}_{-1.5}$ \\ \midrule
 & $\chi^2/\mathrm{dof}$ & 547/396 \\
\midrule
\midrule
\end{tabular}
\end{table}

\subsection{The 0111010101 orbital phase-averaged spectrum}
\label{eclipse}

During the eclipse phases, the direct component from the neutron star is blocked by the counterpart. Thus, X-ray continuum emission is lower than in out of eclipse phases and the emission lines stand out clearly above the continuum level. The strong X-ray radiation from the neutron star ionises the stellar wind producing a plasma with high ionisation parameter \citep[see][for instance]{2003ApJ...582..959W,2005ApJ...634L.161I}. Therefore, the circumstellar gas is expected to be photoionisation dominated and models such as \emph{pion} in \emph{SPEX} \citep{2016A&A...596A..65M} or \emph{photemis} in \emph{XSPEC} are tested. A single absorbed power-law ($\chi_\mathrm{red}^2\sim 721/396$) or a single absorbed black-body ($\chi_\mathrm{red}^2\sim 746/396$) described the continuum reasonably. Then, a photoionised plasma in emission is added to the continuum to describe the emission lines. The ionisation parameter $\log \xi$ (W\,m) is limited between 1 and 4, the covering fraction is set to zero, i.e. pure emission line model, and the solid angle $\Omega$ is set to 1 (a full shell of $4\, \pi$ steradians). A low-energy and high-energy exponential cut-off of the power-law component are applied by using two \emph{etau} components, while the emission from the \emph{pion} component is not attenuated by these two components.

The interstellar absorption was described with the \emph{hot} model in \emph{SPEX}, which calculates the transmission of a collisionally-ionised equilibrium plasma
\citep{2004A&A...423...49D,2005A&A...434..569S}. 
For a given temperature and set of abundances, the model calculates the ionisation balance and determines all the ionic column densities by scaling to the prescribed total hydrogen column density. We fixed the temperature to the minimum value in SPEX (i.e. 0.001 eV) to mimic the neutral interstellar gas and the hydrogen column density to the Galactic absorption toward the direction of Cen X$-$3, $N_H = 1.16\times10^{26}$ m$^{-2}$ \citep{2013MNRAS.431..394W}. We also modelled the intrinsic absorption due to the stellar wind and/or other material around the compact object with a second \emph{hot} component. The best fit parameters of the photoionisation dominated model are tabulated in Table~\ref{tab:ecl-pion}.

\begin{figure}[h!tb]
  \centering
  \includegraphics[angle=-90,width=.99\columnwidth]{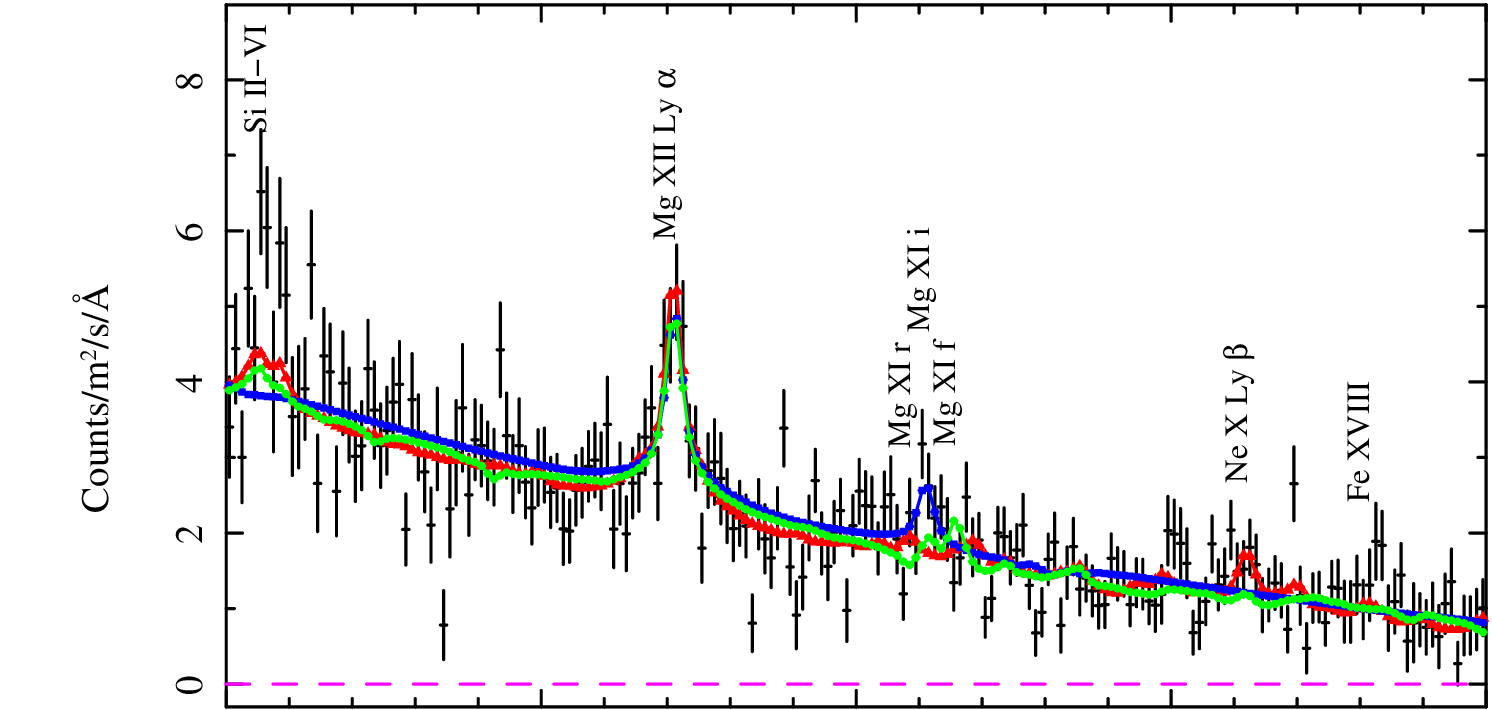}
  \includegraphics[angle=-90,width=1.02\columnwidth]{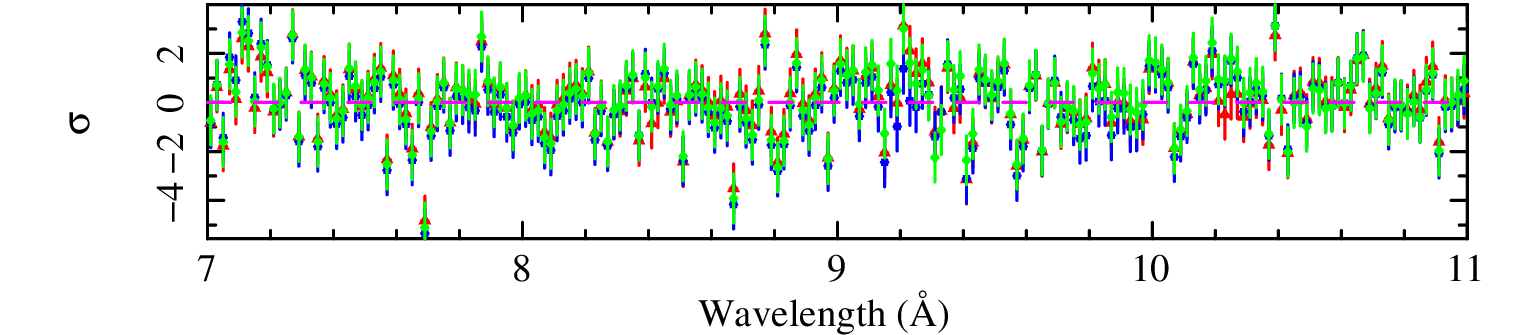}
  \includegraphics[angle=-90,width=.99\columnwidth]{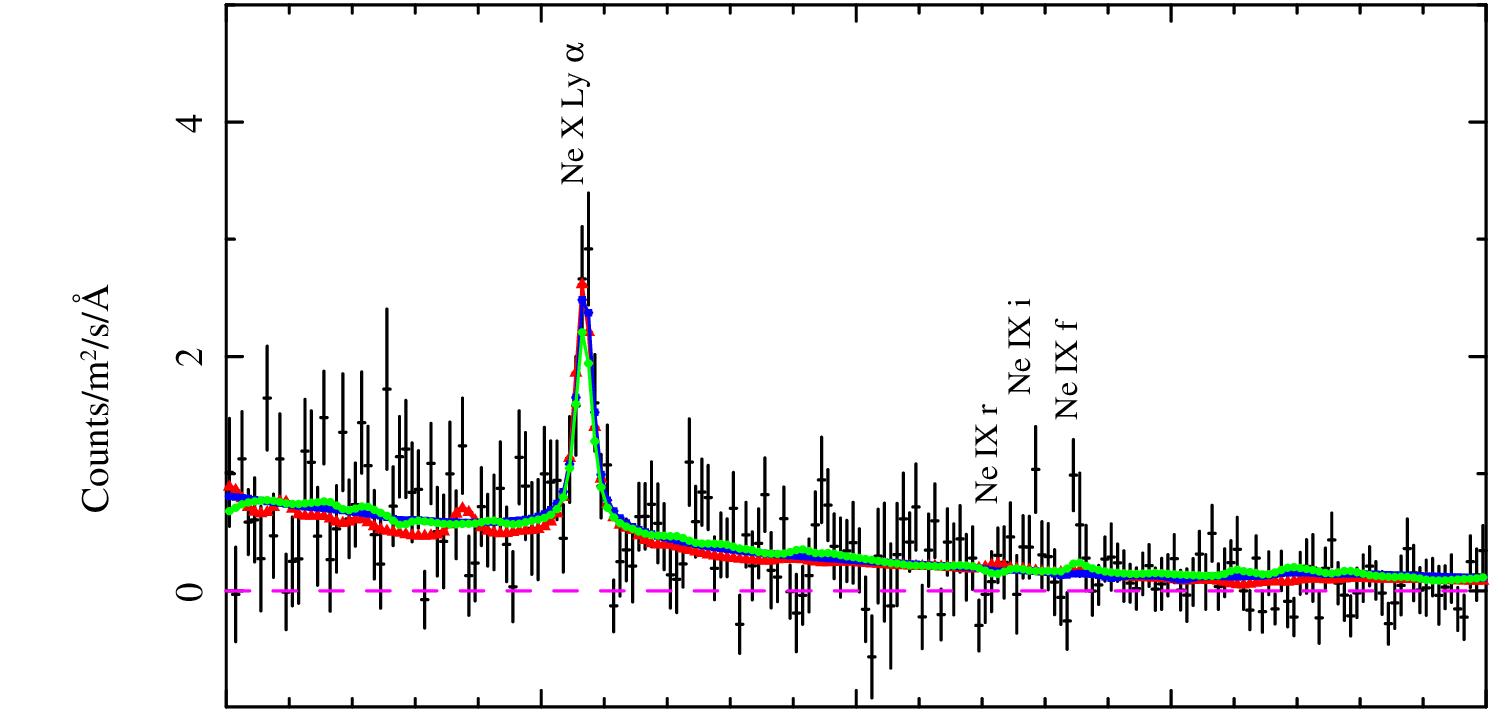}
  \includegraphics[angle=-90,width=1.022\columnwidth]{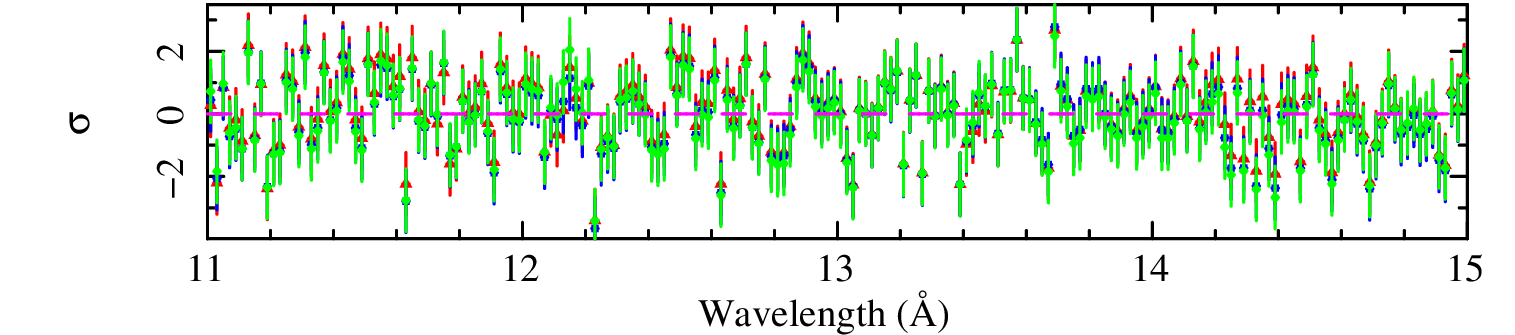}
  \caption{\emph{RGS} data, photoionisation dominated model (green circle), collisional dominated model (red triangle), Gaussian emission lines (blue square) and residuals (expressed in terms of number of standard deviations) for the 0111010101 orbital phase-averaged spectrum.
  }
  \label{fig:ecl-3models}
\end{figure}

As a intrinsic absorber, we use either a \emph{xabs} absorption component or a \emph{warm} absorber instead the second \emph{hot} model. The apparent absorption features in the spectrum are not well described and does not improve the previous fit. It is also tested if the eclipsed spectrum is better described by a plasma in a collisional-ionisation equilibrium state. Therefore, the same continuum components was used but changing the \emph{pion} component by a \emph{cie} component in \emph{SPEX}. This model provides a better description of the data ($\chi^2/\mathrm{dof} = 518/391$) compared to the previous one. Any combination model of a collisionally ionised component and a photoionisation component did not improve these fits.

\begin{table}[h!tb]
\caption{Parameters of the emission lines according to the fit results from the three models.
}
\label{tab:lines}
\centering
\begin{tabular}{lccc}
\midrule
\midrule
 & \multicolumn{3}{c}{Model component} \\
Parameter & \emph{pion} & \emph{cie} & \emph{Gaussian} \\
\midrule
 & \multicolumn{3}{c}{Mg \textsc{XII} Ly $\alpha$} \\ \midrule
\multirow{2}{*}{Flux (10$^{41}$ ph\,s$^{-1}$)} & $7.1$ & $5.9$ & \multirow{2}{*}{$6.8\pm 1.0$} \\
 & $3.6$ & $3.3$ &  \\
\multirow{2}{*}{$\lambda$ (\AA)} & $8.41921$ & $8.41900$ & \multirow{2}{*}{$8.424^{+0.008}_{-0.006}$} \\
 & $8.42462$ & $8.42400$ &  \\
\multirow{2}{*}{FWHM (\AA)} & $0.007$ & $0.005$ & \multirow{2}{*}{$0.005$ (fixed)} \\
 & $0.007$ & $0.005$ &  \\
\midrule
 & \multicolumn{3}{c}{Mg \textsc{XI} (r i f)} \\ \midrule
\multirow{3}{*}{Flux (10$^{41}$ ph\,s$^{-1}$)} & $4.8$ & $1.3$ & \multirow{3}{*}{$3.1\pm 0.9$} \\
 & $1.2$ & $0.2$ &  \\
 & $4.2$ & $0.8$ &  \\
\multirow{3}{*}{$\lambda$ (\AA)} & $9.16875$ & $9.17000$ & \multirow{3}{*}{$9.222^{+0.011}_{-0.013}$} \\
 & $9.23121$ & $9.23200$ &  \\
 & $9.31434$ & $9.31500$ &  \\
\multirow{3}{*}{FWHM (\AA)} & $0.007$ & $0.006$ & \multirow{3}{*}{$0.005$ (fixed)} \\
 & $0.008$ & $0.006$ &  \\
 & $0.008$ & $0.006$ &  \\
\midrule
 & \multicolumn{3}{c}{Ne \textsc{X} Ly $\beta$} \\ \midrule
\multirow{2}{*}{Flux (10$^{41}$ ph\,s$^{-1}$)} & $1.0$ & \multirow{2}{*}{$3.7$} & \multirow{2}{*}{------} \\
 & $1.0$ &  &  \\
\multirow{2}{*}{$\lambda$ (\AA)} & $10.2390$ & \multirow{2}{*}{$10.240$} & \multirow{2}{*}{------} \\
 & $10.2396$ &  &  \\
\multirow{2}{*}{FWHM (\AA)} & $0.008$ & \multirow{2}{*}{$0.007$} & \multirow{2}{*}{------} \\
 & $0.008$ &  &  \\
\midrule
 & \multicolumn{3}{c}{Ne \textsc{X} Ly $\alpha$} \\ \midrule
\multirow{2}{*}{Flux (10$^{42}$ ph\,s$^{-1}$)} & $3.0$ & $2.2$ & \multirow{2}{*}{$2.5\pm 0.3$} \\
 & $1.5$ & $1.2$ &  \\
\multirow{2}{*}{$\lambda$ (\AA)} & $12.1328$ & $12.1320$ & \multirow{2}{*}{$12.140\pm 0.005$} \\
 & $12.1382$ & $12.1375$ &  \\
\multirow{2}{*}{FWHM (\AA)} & $0.010$ & $0.008$ & \multirow{2}{*}{$0.005$ (fixed)} \\
 & $0.010$ & $0.008$ &  \\
\midrule
 & \multicolumn{3}{c}{Ne \textsc{IX} (r i f)} \\ \midrule
\multirow{3}{*}{Flux (10$^{41}$ ph\,s$^{-1}$)} & $8.3$ & $1.4$ & \multirow{3}{*}{------} \\
 & $1.3$ & $0.2$ &  \\
 & $4.9$ & $0.9$ &  \\
\multirow{3}{*}{$\lambda$ (\AA)} & $13.4471$ & $13.4480$ & \multirow{3}{*}{------} \\
 & $13.5529$ & $13.5530$ &  \\
 & $13.6987$ & $13.7000$ &  \\
\multirow{3}{*}{FWHM (\AA)} & $0.011$ & $0.009$ & \multirow{3}{*}{------} \\
 & $0.011$ & $0.009$ &  \\
 & $0.011$ & $0.009$ &  \\
\midrule
\midrule
\end{tabular}
\end{table}

To parametrise the strongest emission lines, the absorbed power-law is also used and the collisionally ionised component or photoionisation component is now substituted by different Gaussian line profiles ($\chi^2/\mathrm{dof} = 550/394$). The full width at half maximum (FWHM $= \sigma\, \sqrt{\ln 256} = 2.3548\, \sigma$) of the Gaussian lines is fixed at $0.005$ \AA\, and the derived line flux is compared to that obtained in the previous models. The strongest features at $8.424$ \AA, (identified as Mg \textsc{xii} which is actually a doublet) and at $12.140$ \AA\, (identified as Ne \textsc{x} which is also a doublet) have a probability of $1.0\times 10 ^{-10}$ and $2.3\times 10^{-7}$, respectively. The faintest emission line at $9.222$ \AA\, (identified as Mg \textsc{xi}, therefore, located in the He-like triplet of this element and unresolved properly in this eclipse spectrum) has a probability of $0.015$. Mg \textsc{xii} Ly $\alpha$ and Ne \textsc{x} Ly $\alpha$ are highly significant at $> 3\, \sigma$ and the wavelengths of the centroid of the lines are equivalent to the reference values. The equivalent width (EW) is the ratio $F_\ell/F_c$, where $F_\ell$ is the photon flux of the line in unit ph\,s$^{-1}$ and $F_c$ is the flux per unit of energy of the continuum at the energy of the line in unit ph\,s$^{-1}$\,keV$^{-1}$. The EW for Mg \textsc{xii} Ly $\alpha$ and for Ne \textsc{x} Ly $\alpha$ are $\sim 19$ eV and $\sim 61$ eV, respectively, consistent with the values found by \citet[][see their Figure~10]{2021MNRAS.501.5892S}. The other emission lines listed in Table~\ref{tab:lines} are detected at $\lesssim 3\, \sigma$.
 Fig.~\ref{fig:ecl-3models} shows the best fit of the eclipse averaged spectrum by the three models. The best-fit values of the emission lines for the three models are shown in Table~\ref{tab:lines}. 

Photoexcitation could play a role in photoionisation equilibrium plasma (\emph{pion}) when multiple photons with the right energy excite the electrons to the resonant level. This process enhances the resonance line and modifies the gas distribution \citep{2021A&A...648A.105A}. Resonance line scattering can also enhance the resonant (\emph{r}) component making it stronger than the intercombination (\emph{i}) or forbidden (\emph{f}) components in the He-like triplet lines. This phenomenon is reported by \citet{2003ApJ...582..959W} for this source, observed during eclipse, and represents a plausible mechanism for increasing the flux of the \emph{r}-line. Nevertheless, it should be noted that the He-like emission lines of Mg \textsc{xi} and Ne \textsc{ix} cannot be well-defined by the models and prevent us to estimate any ratios.

For recombination, as well as emission from thin collisionally ionised plasmas, the two Ly $\alpha$ components are emitted in proportion to their statistical weights, i.e. the flux ratio Ly $\alpha_1$/Ly $\alpha_2 = 2$ and widths of both components are equal \citep{2003ApJ...582..959W}. For Mg \textsc{XII} Ly $\alpha$, the X-ray velocity broadening is about 240 km/s in the \emph{pion} model which is slightly broader than in the \emph{cie} model, 180 km/s. On the other hand, for Ne \textsc{X} Ly $\alpha$, the emission line broadening is 100 km/s and 80 km/s, respectively. Using two Gaussians to describe the Mg Ly $\alpha$ line produced FWHM consistent with zero in both components.

\begin{figure*}[h!tb]
  \centering
  \includegraphics[angle=-90,width=\columnwidth]{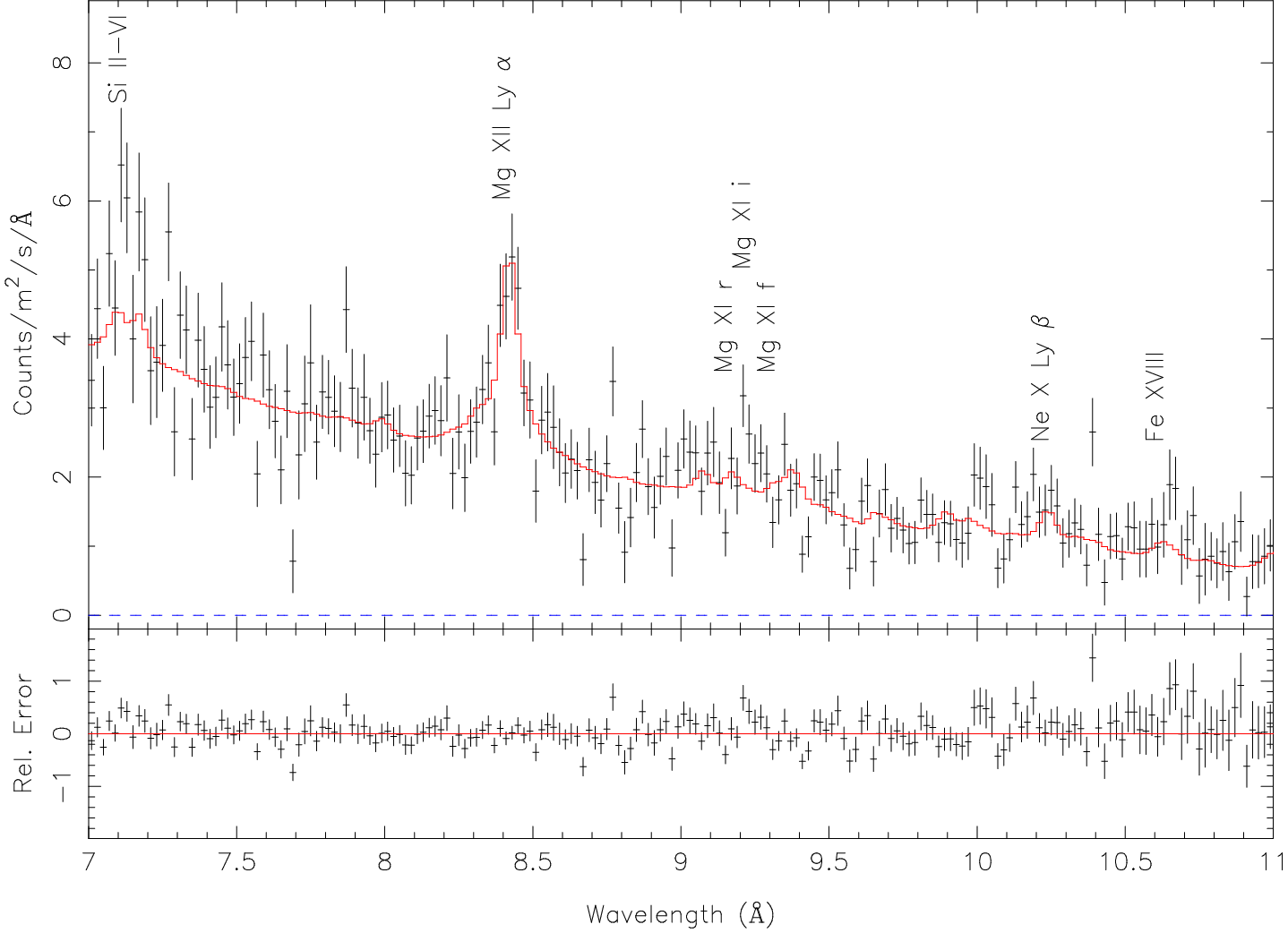}
  \includegraphics[angle=-90,width=\columnwidth]{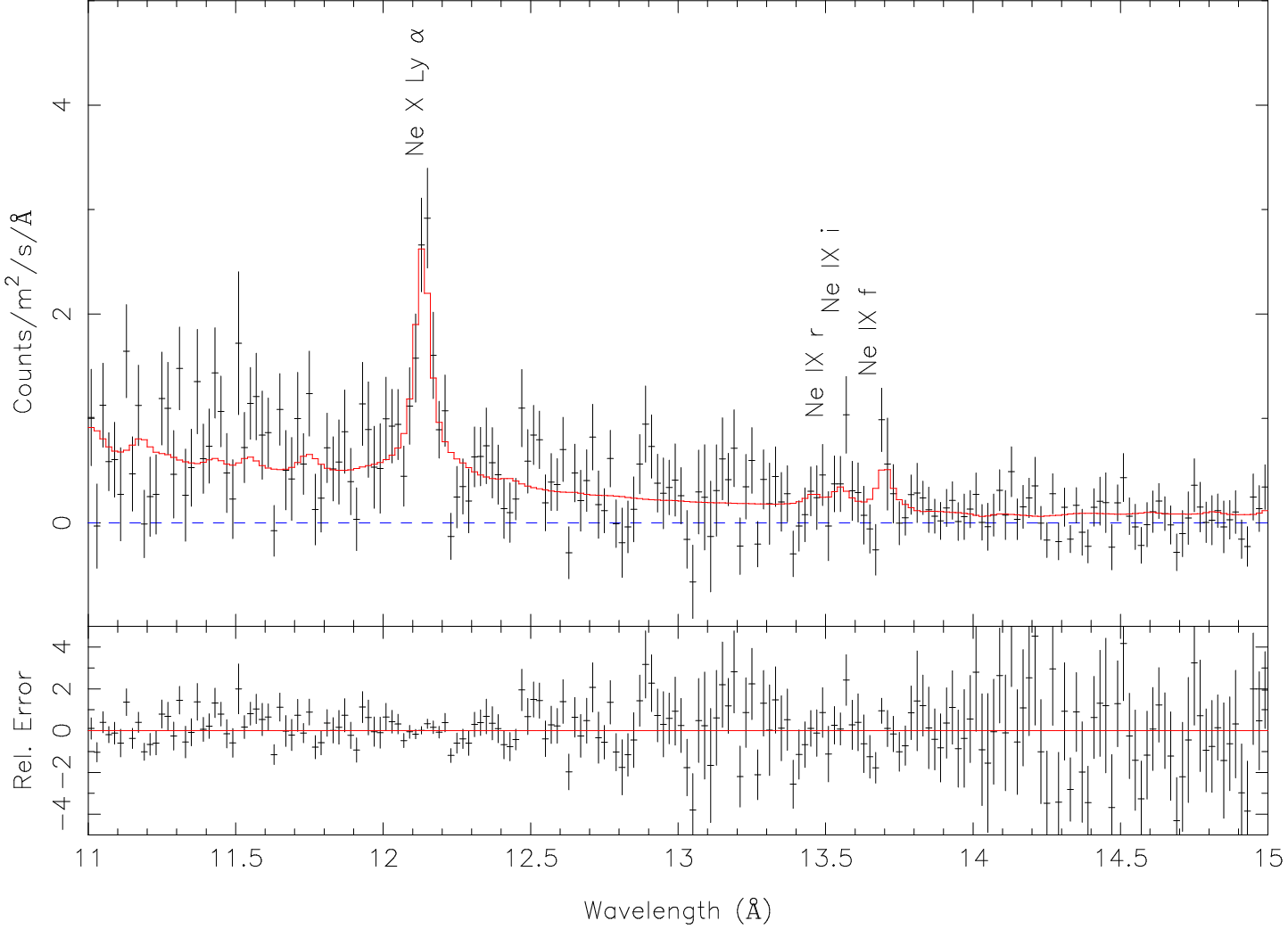}
  \caption{\emph{RGS} data, model (two cie components) and relative residuals for the 0111010101 orbital phase-averaged spectrum.
  }
  \label{fig:ecl-cie}
\end{figure*}

Taking into account that the hydrogen-like features detected in the eclipse spectrum peak at different temperatures, here it is tested if a pure collisional-ionisation equilibrium plasma provides a better description of the data. Therefore, the spectrum is modelled by emission from two CIE models absorbed by neutral gas. One set of abundances, which was free to vary along with the emission measures and temperatures, was adopted. The ion temperature was coupled to the electron temperature as well. The free parameters for the fit are listed in Table~\ref{tab:ecl-2cie}, together with the statistical 1$\sigma$ uncertainties. The X-ray flux is estimated in the energy range (0.83--1.77) keV. This model gives a $\chi^2/dof = 517/391$ and is equivalent to the absorbed power-law and \emph{cie} description. In Fig.~\ref{fig:ecl-cie} we plot the eclipse spectrum of Cen X$-$3 along with the model. Emission lines from Ne, Mg and Si ions are clearly present which are stronger than in the out-of-eclipse where these emission lines cannot be distinguished from the X-ray continuum without doubts (see Sect~\ref{noeclipse}).

\begin{table*}[h!tb]
\caption{Spectral fit parameters to the 0111010101 orbital phase-averaged and phase-resolved spectra.
}
\label{tab:ecl-2cie}
\centering
\begin{tabular}{llcccc}
\midrule
\midrule
Component & Parameter & Averaged & Eclipse & Eclipse-egress & Egress \\
\midrule
\multirow{2}{*}{\emph{Hot$_\mathrm{intrinsic}$}} & $N_H$ (10$^{26}$ m$^{-2}$) & 1.1$\pm$0.5 & 0.47$^{+0.22}_{-0.20}$ & 0.70$\pm 0.24$ & 0.85$^{+0.11}_{-0.16}$ \\
 & $k\,T$ (eV) & 42.0$^{+2.4}_{-1.8}$ & 1$^{+6}_{-1}$ & 34$\pm 16$ & 44$^{+4}_{-2}$ \\ \midrule
\multirow{4}{*}{\emph{Cie}1} & ne nX V (10$^{64}$ m$^{-3}$) & 3.3$^{+0.5}_{-0.6}$ & 1.05$^{+0.15}_{-0.13}$ & 2.2$^{+0.4}_{-0.3}$ & 5.0$^{+0.5}_{-0.4}$ \\
 & $k\,T$ (keV) & 1.8$^{+0.5}_{-0.2}$ & 1.8$^{+0.4}_{-0.2}$ & 2.5$^{+1.4}_{-0.5}$ & 2.29$^{+0.24}_{-0.23}$ \\
 & Flux (10$^{-15}$ W\,m$^{-2}$) & 2.2$^{+0.3}_{-0.4}$ & 0.95$^{+0.14}_{-0.12}$ & 1.9$\pm 0.3$ & 3.8$^{+0.4}_{-0.3}$ \\
 & Unabs. Flux (10$^{-14}$ W\,m$^{-2}$) & 1.7$^{+0.2}_{-0.3}$ & 0.55$^{+0.08}_{-0.07}$ & 1.09$^{+0.19}_{-0.16}$ & 2.46$^{+0.23}_{-0.22}$ \\ \midrule
\multirow{7}{*}{\emph{Cie}2} & ne nX V (10$^{64}$ m$^{-3}$) & 3$^{+7}_{-2}$ & 0.16$^{+0.24}_{-0.15}$ & 0.20$^{+0.24}_{-0.14}$ & 14$^{+9}_{-23}$ \\
 & $k\,T$ (keV) & 0.19$^{+0.11}_{-0.05}$ & 0.42$^{+0.18}_{-0.10}$ & 0.52$^{+0.24}_{-0.20}$ & 0.15$^{+0.04}_{-0.03}$ \\
 & Abundance Ne & 1.5$^{+0.3}_{-0.2}$ & \multicolumn{3}{c}{\emph{Fixed}} \\
 & Abundance Mg & 0.7$^{+0.3}_{-0.1}$ & \multicolumn{3}{c}{\emph{Fixed}} \\
 & Abundance Fe & 0.11$^{+0.06}_{-0.05}$ & \multicolumn{3}{c}{\emph{Fixed}} \\
 & Flux (10$^{-17}$ W\,m$^{-2}$) & 9$^{+24}_{-8}$ & 9$^{+13}_{-7}$ & 10$^{+14}_{-9}$ & 16$^{+24}_{-10}$ \\
 & Unabs. Flux (10$^{-15}$ W\,m$^{-2}$) & 4$^{+9}_{-3}$ & 1.3$^{+1.7}_{-0.9}$ & 1.2$^{+1.7}_{-1.2}$ & 6$^{+10}_{-4}$ \\ \midrule
 & $\chi^2/\mathrm{dof}$ & 517/391 & 386/394 & 383/393 & 475/394 \\
\midrule
\midrule
\end{tabular}
\end{table*}

Thawing abundances of the elements did not produce a significant improvement of the fit. The \emph{slab} model was used to identify other absorbing ions present in the plasma from the line list available in \emph{SPEX} version 2.0 such as Mg \textsc{ix}, Ni \textsc{xxiv}, Ni \textsc{xxii}, Fe \textsc{xxiv}, Fe \textsc{xxiii} or Si \textsc{v}, but the fit did not improve significantly. Changing the electron density of the plasma or the micro-turbulent velocity for line broadening did not improve the description of the data either.

The description of the averaged spectrum with two \emph{cie} components fits well the emission lines of Ne \textsc{x} (12.134 \AA), Fe \textsc{xviii} (10.627 \AA), Ne \textsc{x} (10.240 \AA), Mg \textsc{xii} (8.421 \AA) and Si \textsc{ii-v} (7.126 \AA). If present, the He-like emission lines of Ne \textsc{ix} (13.447 \AA) and Mg \textsc{xi} (9.232 \AA) were too blended to allow the forbidden line and intercombination line to be quantitatively separated and could not be resolved in the eclipse by \emph{RGS}. Assuming that the X-ray emission of the neutron star is blocked nearly completely by the giant counterpart, each \emph{cie} could explain the two different circumsource X-ray emission contributions due to the strong stellar wind (cie2) and from matter around the compact object, i.e. accretion disc or another structure material close to the neutron star (cie1). The temperature of cie2 (stellar wind) is $k\, T=0.19^{+0.11}_{-0.05}$ keV while that of cie1 is much hotter, $k\, T=1.8^{+0.5}_{-0.2}$ keV. Simultaneous \emph{XMM/EPIC} spectra detected the presence of the Fe K$\alpha$ at 6.4 keV, Fe \textsc{xxv} at 6.7 keV and Fe \textsc{xxvi} at 6.97 keV \citep{2012BASI...40..503N,2021MNRAS.501.5892S}. These lines were clearly observed during the eclipse, eclipse-egress and egress periods. Therefore, the ionisation state is expected to be $\xi \sim 10^{3.4}$ erg cm s$^{-1}$ just to detect the helium-like iron line \citep{1996PASJ...48..425E}. Concurrently, to produce Fe K$\alpha$ emission line the ionisation parameter should be $\xi \lesssim 10^{2}$ erg cm s$^{-1}$ and the silicon line intensity ratio [Si \textsc{xiii}/Si \textsc{xiv}] suggests an ionisation parameter in the interval $\left[10^{2.1}-10^{2.4}\right]$ erg cm s$^{-1}$ \citep{2021MNRAS.501.5892S}. This implies that, during the \emph{XMM-Newton} observation, the stellar wind had different levels of ionisation and probably the emission sites may be located in different regions.

\subsection{The 0400550201 orbital phase-averaged spectrum}
\label{noeclipse}

In this orbital phase range we expect to see the direct emission from the neutron star surface and a complex scattered emission due to high-energy electrons and/or the circumsource matter around the compact object. We adopted the same set of abundances derived in the eclipse spectrum. The line-of-sight Galactic absorption towards Cen X$-$3 was frozen using a cold collisionally-ionised component, \emph{hot} in SPEX. Another intrinsic absorption component was added due to any circumsource matter around the pulsar. In all cases, a single continuum component, such as, power law, comptonisation, blackbody or modified blackbody did not give an acceptable fit. Then, several double- or triple-component continuum models were tested. A combination of a power-law and a blackbody components gave a $\chi^2/dof = 1625/646$. According to this result, the photon index $\Gamma$ of the spectrum was $\sim -0.7$, the emitting region was at a blackbody temperature of $\sim 0.2$ keV and the blackbody radius was found to be $\sim 110$ km, which is inconsistent with an entire neutron star surface emission. Replacing the blackbody by an inverse Comptonisation of soft photons in a hot plasma (\emph{comt}) model led worst results ($\chi^2/dof = 1814/645$). In this case,  the photon index $\Gamma$ of the spectrum was $\sim 2.2$ and the plasma temperature $t_1$ was $\sim 900$ keV, larger than the upper limit where this model could be applied and other processes such as a pair production should be take into account in modelling the emitted radiation \citep{1995ApJ...449..188H}.

Fitting two or more \emph{cie} components did not describe the out-of-eclipse spectrum satisfactorily. Next we tried a continuum component modelled with a power-law (direct emission from the NS) and a \emph{cie} component absorbed by photoionised gas close to the neutron star (\emph{xabs}1), partially covered photoionised gas far away the X-ray source (\emph{xabs}2) and interstellar medium absorption (\emph{hot}). The fit provides a $\chi^2/dof = 1359/643$ (dof means degrees of freedom), which is poor.
The power-law component was changed by a modified blackbody obtaining a $\chi^2/dof = 1210/641$. Adding a new component did not produce any significant improvement.

For this observation, simultaneous \emph{XMM/EPIC} spectra detected the presence of the Fe K$\alpha$ at $\sim$6.4 keV, Fe \textsc{xxv} at 6.7 keV and Fe \textsc{xxvi} at 6.97 keV \citep{2021MNRAS.501.5892S} implying that the stellar wind required an ionisation parameter $\xi \sim 10^{3.4}$ erg\,cm\,s$^{-1}$ \citep{1996PASJ...48..425E}. This spectrum may have both emission and absorption features superimposed on the X-ray continuum which was described by a power-law plus a comptonisation components. The emission lines were described by two \emph{pion} components defined in the same way as in Sect.~\ref{eclipse}. Although the \emph{pion} model is capable of modelling both of them, for the absorbing gas a \emph{xabs} or \emph{slab} model was used in order to save substantial computational time. In this case, photons from both comptonisation and power-law components are the photoionising source of the \emph{pion} emission components and screened by the absorber component as well as the galactic absorption before reaching the detector. The spectrum is not well described with a $\chi^2/dof = 1519/639$.

\begin{figure}[h!tb]
   \includegraphics[angle=-90,width=.97\columnwidth]{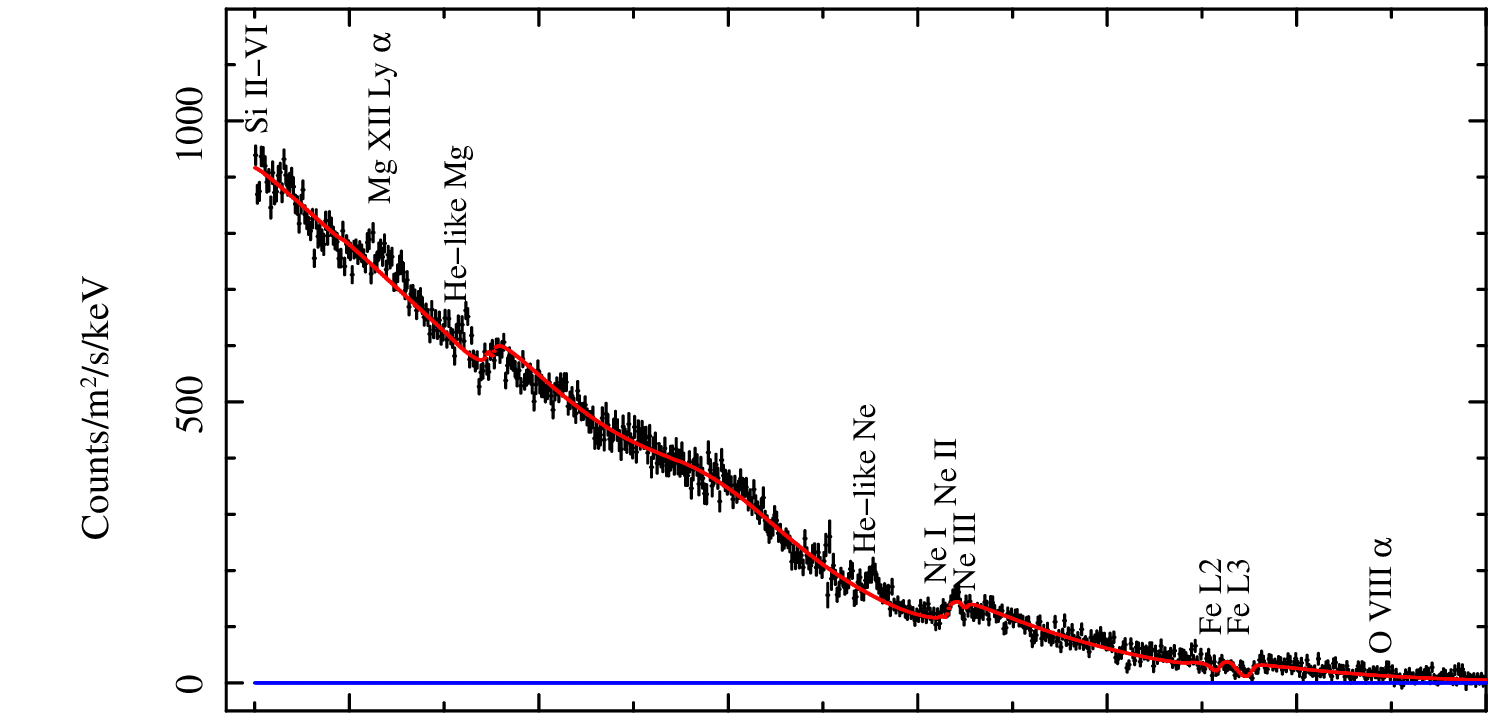}
 \includegraphics[angle=-90,width=\columnwidth]{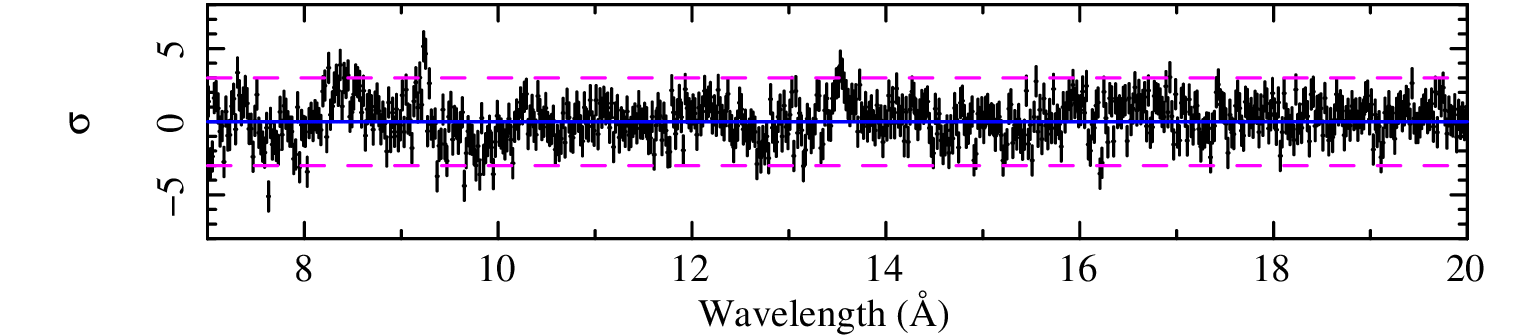}
 
  \caption{\emph{RGS} data, continuum model and residuals (in terms of number of standard deviations) for the 0400550201 orbital phase-averaged spectrum. The dashed lines indicate possible emission/absorption lines detected at $3\, \sigma$ level.
  }
  \label{fig:noecl}
\end{figure}

Finally, the spectral continuum is described by two absorbed power-laws plus a broad Gaussian component for fitting the significant deviations around 12 \AA. This fit provides a reasonable description of the continuum X-ray emission with a $\chi^2/dof = 1112/642$. The free parameters for the best fit are listed in Table~\ref{tab:noecl}. The X-ray flux is estimated in the energy range (0.62--1.77) keV. The corresponding spectrum, continuum model and residuals are presented in Fig.~\ref{fig:noecl}. From the residuals it can be seen that several emission lines are detected at $3\, \sigma$ but the absorption features, if present, cannot be detected significantly (see dashed lines in Fig.~\ref{fig:noecl}, bottom panel). Therefore, the continuum parameters were fixed and the prominent Helium-like lines (Mg \textsc{xi} and Ne \textsc{ix}) and Hydrogen-like lines (Mg \textsc{xii} and Ne \textsc{x}) are fitted with Gaussians to obtain their properties. The width of the lines was fixed to 0.005 \AA\, except for the He-like Ne line which was a free parameter in the fit. The EW for Mg \textsc{xii} Ly $\alpha$ is $\sim 5.8$ eV but the centre of the line is blue-shifted with respect to the reference position. On the other hand, the high level of the continuum around 12 \AA\, did not allow us to detect the Ne \textsc{x} Ly $\alpha$ line. The He-like Mg is not well-resolved in the out-of-eclipse averaged spectrum but the centre of the Gaussian profile points out to a dominant intercombination line compared to the resonance and forbidden components. \citet{2021MNRAS.501.5892S} reported the detection of the He-like Mg and showed a strong intercombination line in the \emph{Chandra} observation of this source (see their Fig.~12). All the best-fit values of the emission features are reported in Table~\ref{tab:gaus}.

\begin{table*}[h!tb]
\caption{Spectral fit parameters to the 0400550201 orbital phase-averaged spectrum, high count rate (HCR) spectrum and low count rate (LCR) spectrum and the corresponding dips and out-of-dips as defined in Fig.~\ref{fig:dips-nodips}. Flux and unabsorbed flux are in units of W\,m$^{-2}$ and Norm of power-law components in units of ph\,s$^{-1}$\,keV$^{-1}$.
}
\label{tab:noecl}
\centering
\begin{tabular}{llccccccc}
\midrule
\midrule
Model & Parameter & Averaged & HCR & dips & no-dips & LCR & dips & no-dips \\
\midrule
\multirow{2}{*}{\emph{hot}} & $N_H$ (10$^{26}$ m$^{-2}$) & 6.0$^{+0.4}_{-0.3}$ & 5.4$\pm$0.4 & 3.5$^{+0.5}_{-0.4}$ & 5.5$^{+0.5}_{-0.4}$ & 8.5$^{+1.2}_{-1.1}$ & 7.8$^{+1.4}_{-1.1}$ & 8.1$^{+1.4}_{-1.2}$ \\
 & $t$ (eV) & 56.0$^{+1.9}_{-1.8}$ & 55.3$\pm$2.3 & 51$\pm 4$ & 55.6$\pm 2.4$ & 59$\pm$5 & 54$^{+8}_{-6}$ & 59$\pm$5 \\
\midrule
\multirow{4}{*}{\emph{PL$_\mathrm{1}$}} & Norm (10$^{44}$) & 122$^{+18}_{-15}$ & 95$^{+17}_{-13}$ & 80$^{+40}_{-30}$ & 96$^{+18}_{-14}$ & 190$^{+130}_{-70}$ & 330$^{+240}_{-120}$ & 170$^{+120}_{-70}$ \\
 & $\Gamma_1 = \Gamma_2$ & 1.48$\pm$0.24 & 1.44$\pm$0.03 & 1.8$^{+0.2}_{-0.3}$ & 1.43$\pm 0.03$ & 1.20$\pm$0.05 & 1.74$\pm 0.12$ & 1.21$\pm$0.05 \\
 & Flux (10$^{-14}$) & 1.65$^{+0.24}_{-0.20}$ & 1.8$\pm$0.3 & 3.8$^{+2.0}_{-1.6}$ & 1.8$\pm 0.3$ & 0.8$^{+0.5}_{-0.3}$ & 1.4$^{+1.1}_{-0.5}$ & 0.8$^{+0.6}_{-0.3}$ \\
 & Unabs. Flux (10$^{-12}$) & 2.8$^{+0.4}_{-0.3}$ & 2.2$^{+0.4}_{-0.3}$ & 1.8$^{+1.0}_{-0.7}$ & 2.2$^{+0.4}_{-0.3}$ & 4$^{+3}_{-2}$ & 7$^{+5}_{-3}$ & 4$^{+3}_{-2}$ \\
\midrule
\multirow{3}{*}{\emph{PL$_\mathrm{2}$}} & Norm (10$^{44}$) & 20.16$^{+0.12}_{-0.13}$ & 20.94$^{+0.18}_{-0.19}$ & 15.6$^{+1.2}_{-1.0}$ & 21.20$^{+0.17}_{-0.18}$ & 16.73$^{+0.18}_{-0.20}$ & 14.6$\pm 0.3$ & 17.30$^{+0.21}_{-0.22}$ \\
 & Flux (10$^{-14}$) & 10.21$\pm$0.06 & 10.75$^{+0.09}_{-0.10}$ & 7.1$\pm 0.5$ & 10.94$\pm 0.09$ & 9.34$^{+0.10}_{-0.11}$ & 6.77$\pm 0.13$ & 9.65$\pm 0.12$ \\
 & Unabs. Flux (10$^{-13}$) & 4.59$\pm$0.03 & 4.79$\pm$0.04 & 3.5$^{+0.3}_{-0.2}$ & 4.86$\pm 0.04$ & 3.93$^{+0.04}_{-0.05}$ & 3.25$\pm 0.06$ & 4.06$\pm 0.05$ \\
 \midrule
\multirow{5}{*}{\emph{Gaussian}} & Norm (10$^{44}$ ph\,s$^{-1}$) & 0.50$\pm$0.05 & 0.54$\pm$0.07 & 0.7$^{+0.5}_{-0.4}$ & 0.56$\pm 0.06$ & 0.59$^{+0.09}_{-0.08}$ & \multirow{5}{*}{------} & 0.58$^{+0.10}_{-0.09}$ \\
 & $\lambda$ (\AA) & 12.12$\pm$0.05 & 12.07$^{+0.06}_{-0.05}$ & 11.5$^{+0.5}_{-0.4}$ & 12.04$^{+0.05}_{-0.04}$ & 12.22$^{+0.09}_{-0.08}$ &  & 12.25$\pm 0.09$ \\
 & FWHM (\AA) & 1.48$^{+0.18}_{-0.17}$ & 1.44$^{+0.24}_{-0.23}$ & 3.6$^{+1.3}_{-1.1}$ & 1.38$^{+0.20}_{-0.18}$ & 1.8$^{+0.3}_{-0.2}$ &  & 1.8$\pm 0.3$ \\
 & Flux (10$^{-15}$) & 1.21$^{+0.13}_{-0.11}$ & 1.34$^{+0.18}_{-0.17}$ & 2.8$^{+2.0}_{-1.4}$ & 1.41$^{+0.16}_{-0.15}$ & 1.37$^{+0.21}_{-0.18}$ &  & 1.31$^{+0.23}_{-0.20}$ \\
 & Unabs. Flux (10$^{-14}$) & 1.07$^{+0.11}_{-0.10}$ & 1.17$^{+0.16}_{-0.14}$ & 1.6$^{+1.2}_{-0.8}$ & 1.21$^{+0.14}_{-0.13}$ & 1.25$^{+0.19}_{-0.17}$ &  & 1.22$^{+0.21}_{-0.18}$ \\
 \midrule
 & $\chi^2/\mathrm{dof}$ & 1112/642 & 1046/642 & 744/627 & 1036/642 & 801/642 & 674/617 & 799/642 \\
\midrule
\midrule
\end{tabular}
\end{table*}

\begin{table}
\caption{He-like and H-like emission lines detected at 3 $\sigma$ level. FWHM was fixed to $0.005$ \AA\, except for the He-like Ne. In the LCR spectrum no emission lines, if present, are detected at 3 $\sigma$ level.
}
\label{tab:gaus}
\centering
\begin{tabular}{llcc}
\midrule
\midrule
\multirow{2}{*}{Line} & \multirow{2}{*}{$\lambda_\mathrm{obs}$ (\AA)} & Flux & \multirow{2}{*}{FWHM (\AA)} \\
 &  & ($\mathrm{ph}\, \mathrm{m}^{-2}\, \mathrm{s}^{-1}$) &  \\
\midrule
 & \multicolumn{3}{c}{\emph{Averaged spectrum}} \\
 \midrule
He-like Ne & 13.541$^{+0.007}_{-0.009}$ & 0.52$^{+0.08}_{-0.07}$ & 0.07$^{+0.04}_{-0.03}$ \\
He-like Mg & 9.233$^{+0.007}_{-0.006}$ & 0.76$\pm$0.14 & $0.005$ \\
H-like Mg & 8.360$^{+0.014}_{-0.022}$ & 0.43$\pm$0.19 & $0.005$ \\
\midrule
 & \multicolumn{3}{c}{\emph{HCR spectrum}} \\
 \midrule
He-like Ne & 13.536$^{+0.007}_{-0.009}$ & 0.62$\pm 0.21$ & 0.05$^{+0.04}_{-0.05}$ \\
H-like Ne & 10.257$^{+0.012}_{-0.013}$ & 0.57$^{+0.15}_{-0.16}$ & $0.005$ \\
He-like Mg & 9.233$^{+0.010}_{-0.005}$ & 0.88$\pm$0.18 & $0.005$ \\
H-like Mg & 8.364$^{+0.015}_{-0.012}$ & 0.83$\pm$0.23 & $0.005$ \\
\midrule
 & \multicolumn{3}{c}{\emph{HCR out-of-dips spectrum}} \\
 \midrule
He-like Ne & 13.541$^{+0.008}_{-0.009}$ & 0.6$^{+1.9}_{-0.3}$ & 0.01$^{+0.03}_{-0.01}$ \\
H-like Ne & 10.253$^{+0.014}_{-0.010}$ & 0.58$\pm 0.17$ & $0.005$ \\
He-like Mg & 9.233$^{+0.010}_{-0.005}$ & 0.96$^{+0.21}_{-0.19}$ & $0.005$ \\
H-like Mg & 8.364$^{+0.012}_{-0.013}$ & 0.96$\pm$0.24 & $0.005$ \\
\midrule
\midrule
\end{tabular}
\end{table}

A visual inspection of the residuals suggests the presence of an absorption feature around 9.4 \AA, consistent with the location of the Be-like Mg \textsc{ix} K$\alpha$ at 9.378 \AA\, \citep{2017A&A...608A.143G} and could be affected by the He-like Mg \textsc{xi} triplet which is no resolved/detected by \emph{RGS}. Under astrophysical conditions, \citet{2015A&A...579A..87M} showed that He-like triplet lines of a given element can be affected by absorption from Li-like ions in the same medium. Although they discussed this effect for the cases of iron and oxygen in both CIE and PIE plasmas, it could be extended to other elements. However, the simultaneous presence of the H-like and He-like iron, which requires a high parameter of ionisation, indicates that this absorption may be due to the He-like Mg \textsc{xi} K$\alpha$ or an overlap with Ne \textsc{ix} rather than Mg \textsc{ix}. Moreover, the line significance is $\lesssim 3\, \sigma$ and the inclusion of an \emph{slab} component did not improve the fit significantly, therefore, this component is not used further.

 Other emission and absorption lines present in out-of-eclipse are complicated to distinguish from the X-ray continuum emission at a $3\, \sigma$ detection level. Apparently, the spectrum also shows the neutral absorption edges of neon and iron. Near to the Ne \textsc{i} edge at 14.3 \AA\, two weak absorption features due to mildly ionised Ne \textsc{ii-iii} could be visible too. The iron L2 and L3 edges are located at 17.2 and 17.5 \AA. Another weak emission line at $\sim$19.0 \AA\, presumably due to O \textsc{viii} $\alpha$, could also be present.

\subsection*{He-like triplet of Ne}
\label{henetriplet}

\begin{figure}[htb]
  \centering
  \includegraphics[angle=-90,width=\columnwidth]{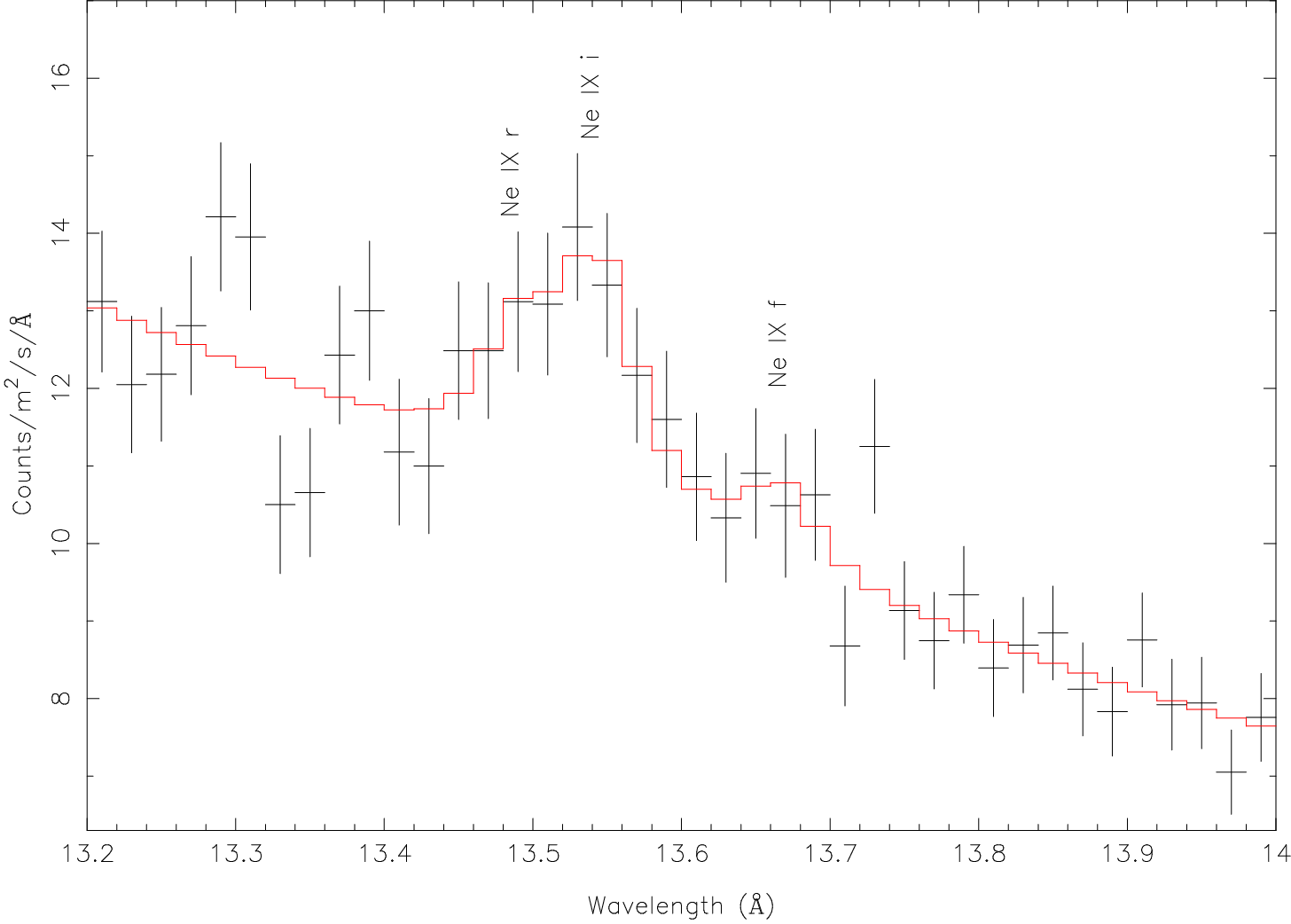}
  \caption{The Ne \textsc{ix} triplet with best fit model. Resonance (r), intercombination (i) and forbidden (f) components are indicated at observed wavelengths.
  }
  \label{fig:ne3}
\end{figure}

The average spectrum showed evidence for the unresolved triplet of Ne \textsc{ix} where the forbidden line is nearly suppressed. For a high density plasma there is an escape channel for the first excited level, upper level of the forbidden line ($f$), by means of collisions. In this case, it is easy to depopulate this state in favour of the upper levels of the intercombination line ($i$), hence implying a reduction of the ratio $\mathcal{R} = f/i$. Nevertheless, the real reason for the low $f/i$ ratio in the spectrum of O-stars and B-stars is the strong ultraviolet (UV) radiation field. In this case, the ion absorbs an UV photon depopulating the forbidden line in favour of the intercombination line \citep{2001A&A...376.1113P}. Therefore, in order to estimate line parameters, it was fitted a local continuum around the Ne \textsc{ix} He-like triplet at $\sim$13.5 \AA\, adding delta line components for modelling this complex feature (Fig.~\ref{fig:ne3}). The parameters are given in Table~\ref{tab:ne}.

Using the values quoted in Table~\ref{tab:ne} it is obtained $\mathcal{R}_\mathrm{obs} = f/i = 0.3^{+0.4}_{-0.3}$, with an upper limit of $\mathcal{R}_\mathrm{obs}<0.7$. This shows that UV photospheric field should be important at the line production site according to \citet{2000A&AS..143..495P}. Since the intensity of the UV radiation decreases outwards in proportion to the dilution factor, i.e. $W(r) = \left[ 1 - \left(1 - (R_\star/r)^2 \right)^{(1/2)} \right]/2$ \citep{mewe1978}, the observed $f/i$ ratio allowed us to estimate the radial location of the Ne He-like triplet \citep[e.g. see Fig. 4 in][]{raassen2008}. The upper limit value for formation radius corresponds to $\lesssim$8 stellar radii and the lower limit should be $\gtrsim$1.5 stellar radii derived from several studies of the hot plasma for O-stars \citep[e.g. see][]{waldron2001,leutenegger2006}. On the other hand, when radiation field is the dominant effect $\mathcal{R}$ becomes a function of the ionisation fraction, electron density $n_e$ and electron temperature $T_e$. Therefore, assuming the peak Ne \textsc{ix} temperature of $T_e \sim 2$ MK and \citet{2000A&AS..143..495P} calculated values (see their Figure 8) for $f/i$ ratios the electron density should be $\gtrsim 10^{12}$ cm$^{-3}$ which is supported by the observed ratio $\mathcal{G}_\mathrm{obs}=(f+i)/r=2.5\pm2.4$ \citep[see Figure 7 in][]{2000A&AS..143..495P}. Thus, these results point out to a hybrid plasma where the effect of collisional plasma is non negligible and probably located inside the Roche lobe radius of Cen X$-$3 \citep{2021MNRAS.501.5892S}.

\begin{table}
\caption{Neon He-like triplet in the out-of-eclipse and HCR spectra.}
\label{tab:ne}
\centering
\begin{tabular}{llcc}
\midrule
\midrule
 & Ion & $\lambda_\mathrm{obs}$ (\AA\,) & Flux ($\mathrm{ph}\, \mathrm{m}^{-2}\, \mathrm{s}^{-1}$) \\
\midrule
\multirow{3}{*}{\emph{Averaged}} & Ne \textsc{ix} r & 13.49$^{+0.02}_{-0.03}$ & 0.17$\pm$0.12 \\
 & Ne \textsc{ix} i & 13.544$^{+0.014}_{-0.012}$ & 0.32$\pm$0.11 \\
 & Ne \textsc{ix} f & 13.67$^{+0.02}_{-0.03}$ & 0.11$^{+0.07}_{-0.06}$ \\
\midrule
\multirow{3}{*}{\emph{HCR}} & Ne \textsc{ix} r & 13.477$^{+0.022}_{-0.017}$ & 0.16$^{+0.12}_{-0.10}$ \\
 & Ne \textsc{ix} i & 13.536$^{+0.010}_{-0.007}$ & 0.47$^{+0.10}_{-0.11}$ \\
 & Ne \textsc{ix} f & 13.69$\pm 0.04$ & 0.11$^{+0.06}_{-0.08}$ \\
\midrule
\midrule
\end{tabular}
\end{table}

Unfortunately, uncertainties of $\mathcal{R}_\mathrm{obs}$ and $\mathcal{G}_\mathrm{obs}$ are quite large and no other He-like lines were resolved neither in the \emph{RGS} spectrum nor in the simultaneous \emph{EPIC} spectra to confirm this argument. On the other hand, thanks to the high energy resolution of \emph{Chandra} \citet{2005ApJ...634L.161I} resolved the He-like Fe \textsc{xxv} and derived ratios $\mathcal{R}_\mathrm{obs} = 0.7\pm 0.4$ and $\mathcal{G}_\mathrm{obs} = 3.1^{+2.2}_{-2.3}$ compatible with \emph{XMM-Newton} values in this work. They also noted that the H-like Fe \textsc{xxvi} was not detected so the stellar wind had a lower ionisation parameter than in both \emph{ASCA} \citep{1996PASJ...48..425E} and \emph{XMM-Newton} \citep{2021MNRAS.501.5892S} observations. More recently, \citet{2024A&A...690A.360S} studied a target of opportunity (ToO) \emph{Chandra} observation during orbital phase (0.13--0.59). They reported the detection of some He-like triplet lines in the spectra of Cen X$-$3 and could determine the parameters $\mathcal{R}_\mathrm{obs}$ and $\mathcal{G}_\mathrm{obs}$ of the Mg \textsc{xi} and Si \textsc{xiii} ions \citep[see Table~6 in][]{2024A&A...690A.360S} which are located in the \emph{RGS} energy range. Their results are broadly consistent with that obtained in this work, when taking the associated errors into account.

\subsection{Orbital phase-resolved spectroscopy}
\label{orbitspec}

\begin{figure*}[h!tb]
  \centering
  \includegraphics[angle=-90,width=\columnwidth]{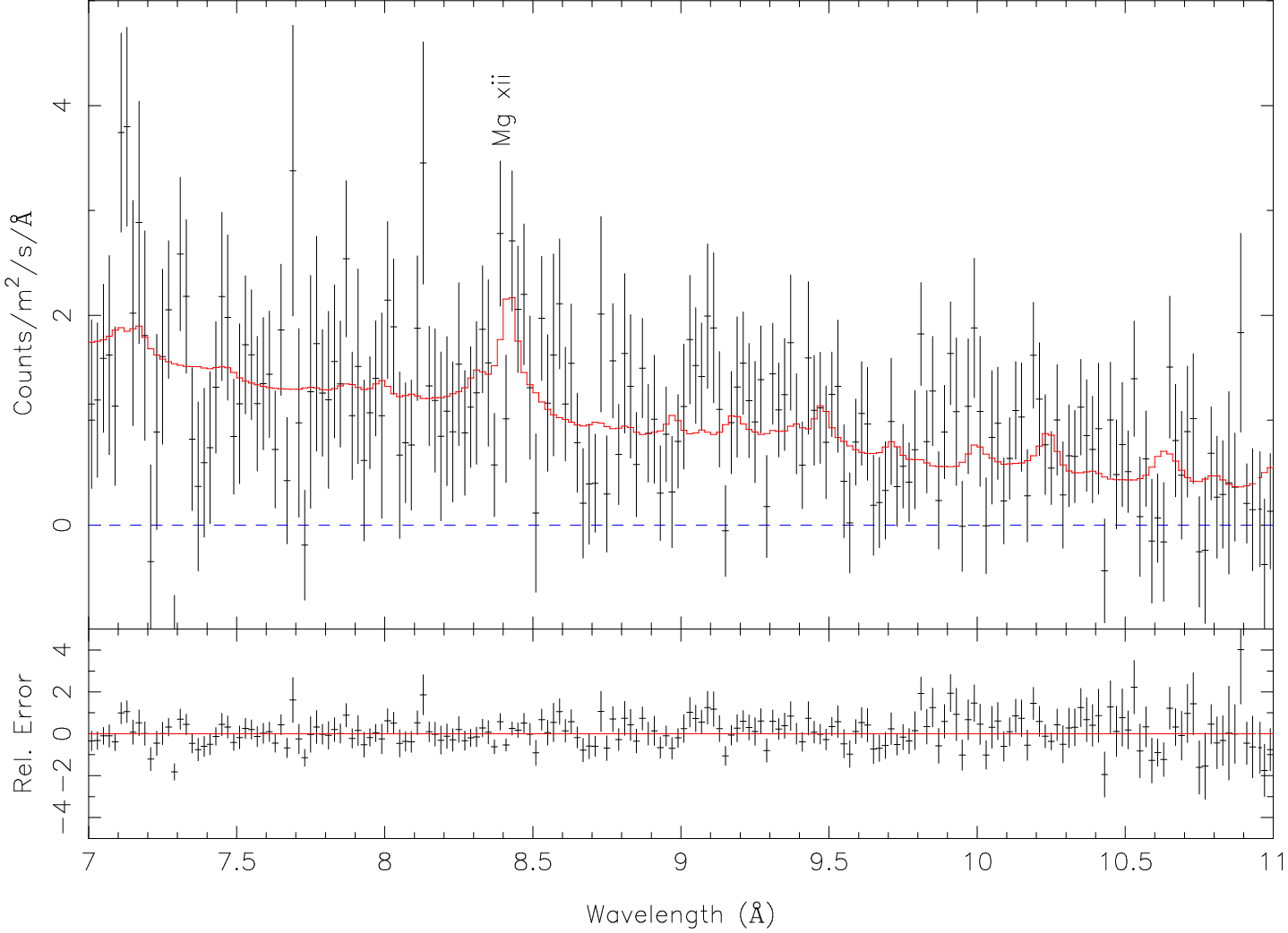}
  \includegraphics[angle=-90,width=\columnwidth]{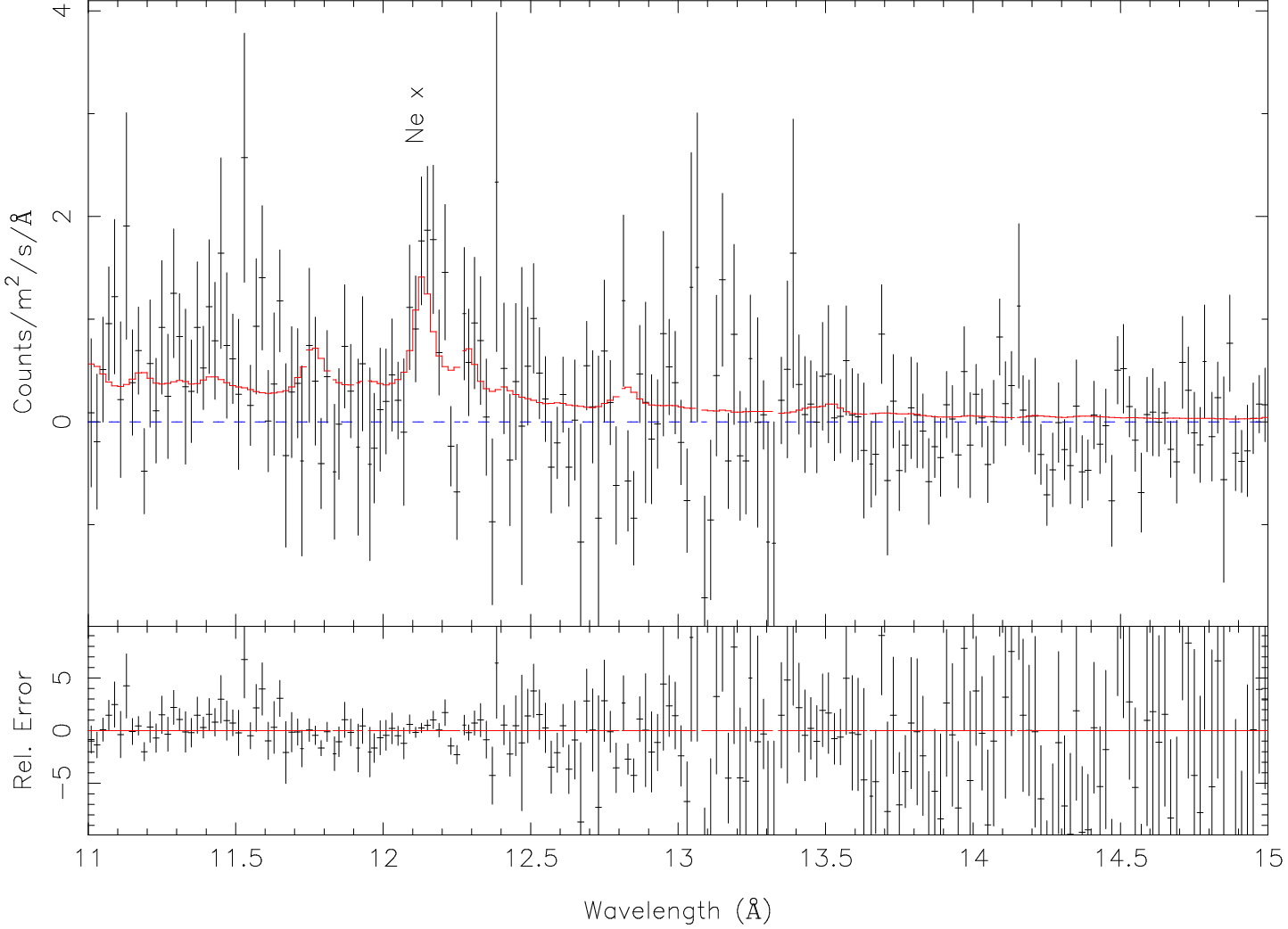}
  \includegraphics[angle=-90,width=\columnwidth]{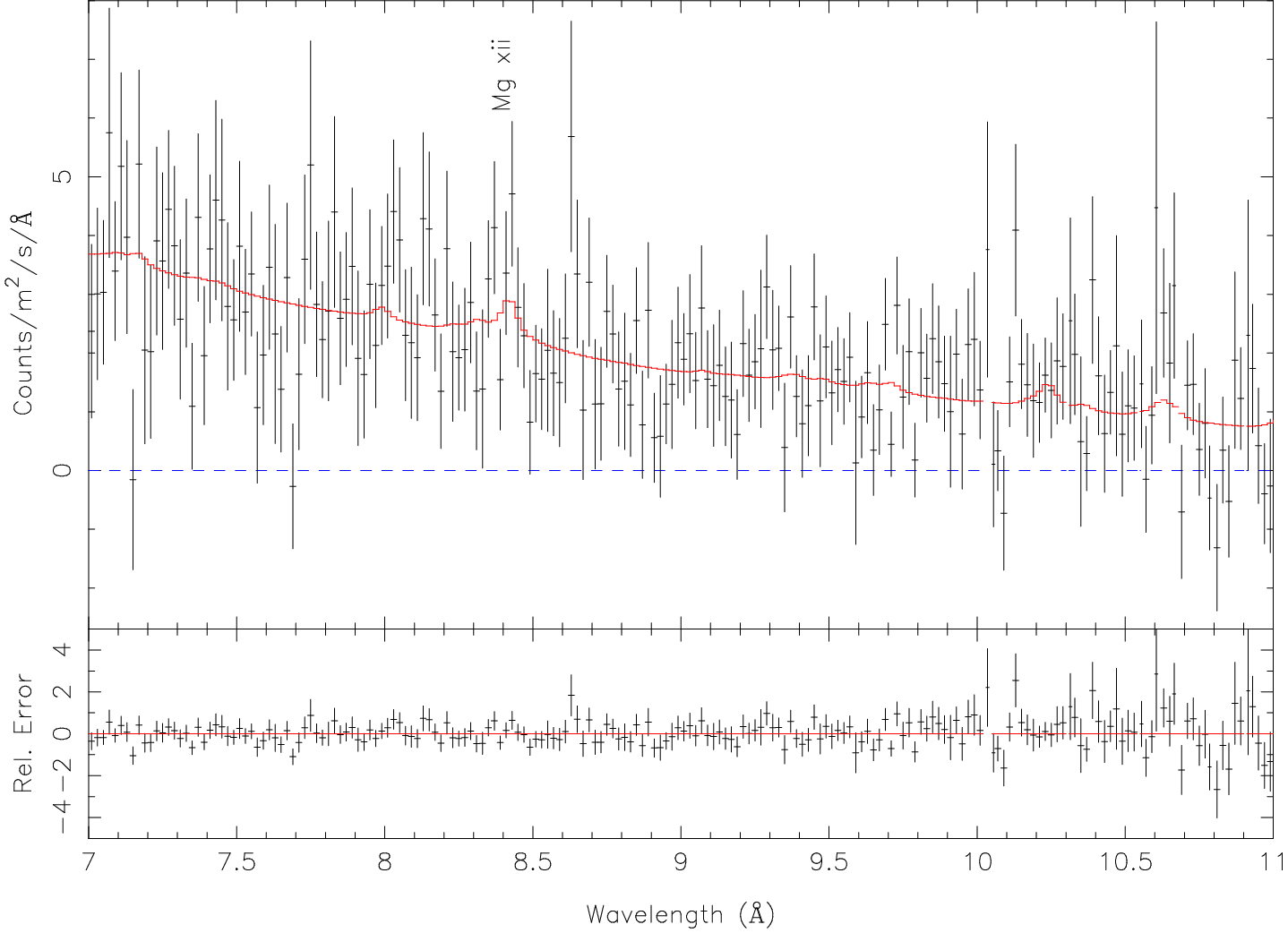}
  \includegraphics[angle=-90,width=\columnwidth]{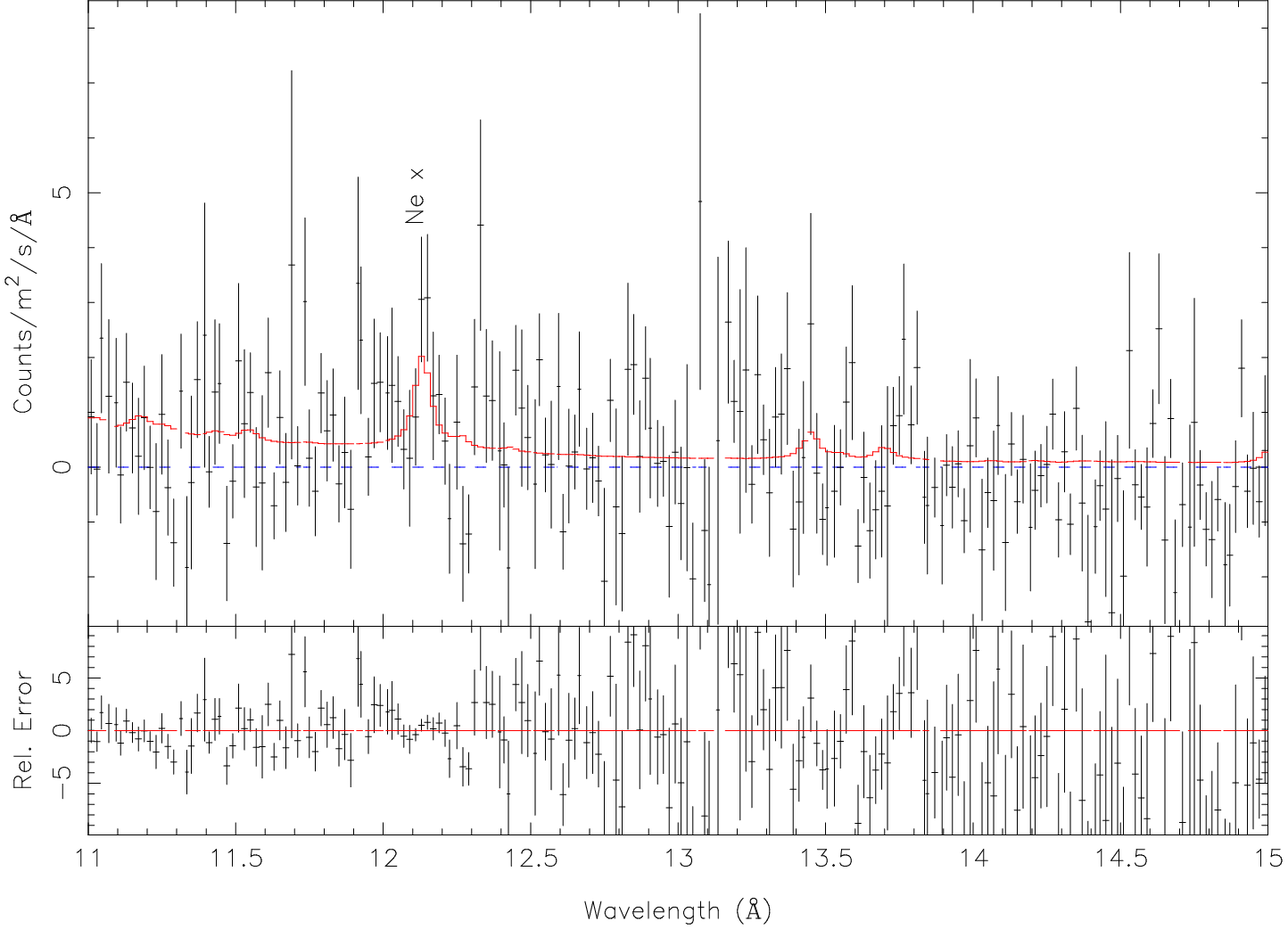}
  \includegraphics[angle=-90,width=\columnwidth]{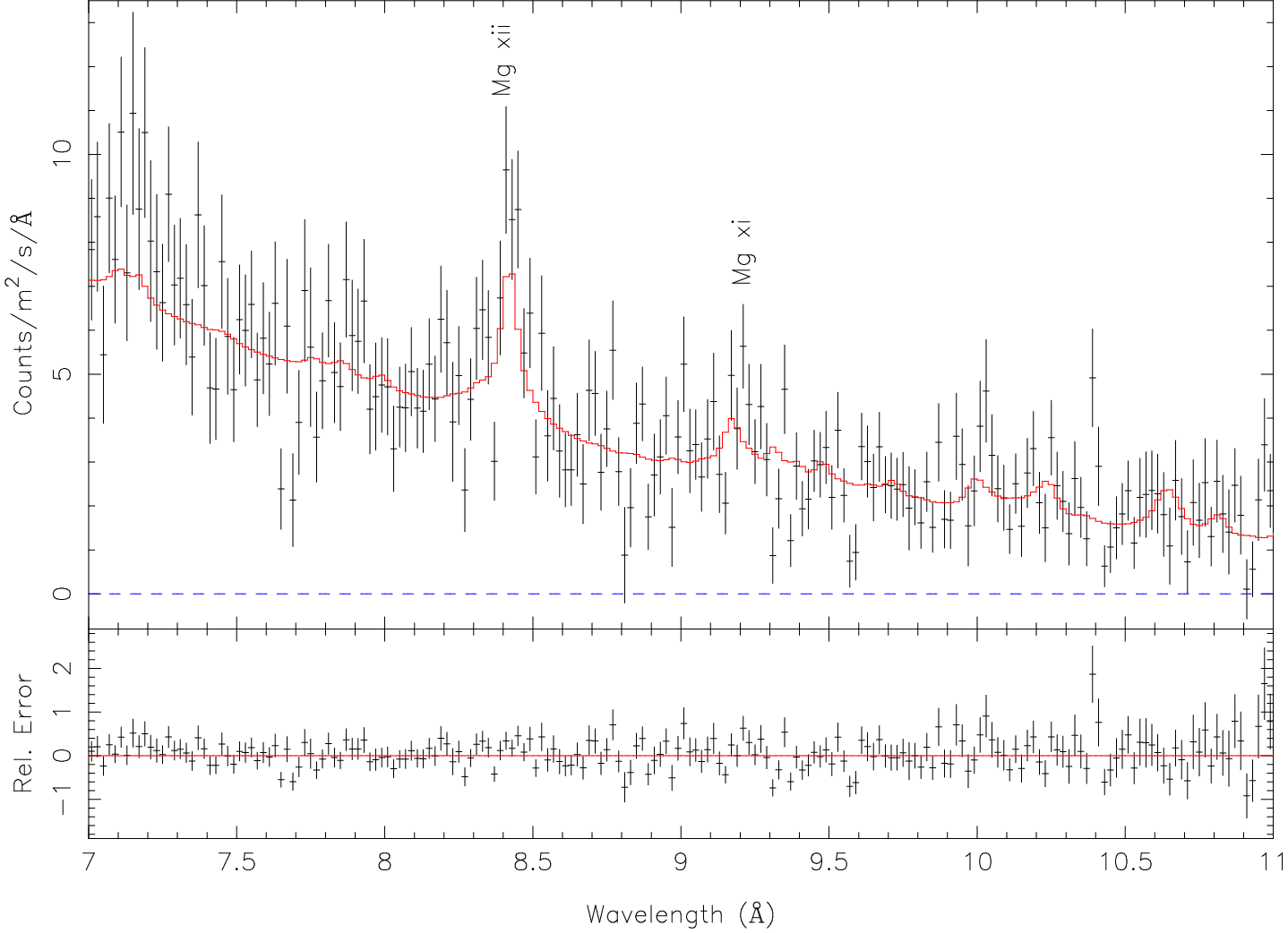}
  \includegraphics[angle=-90,width=\columnwidth]{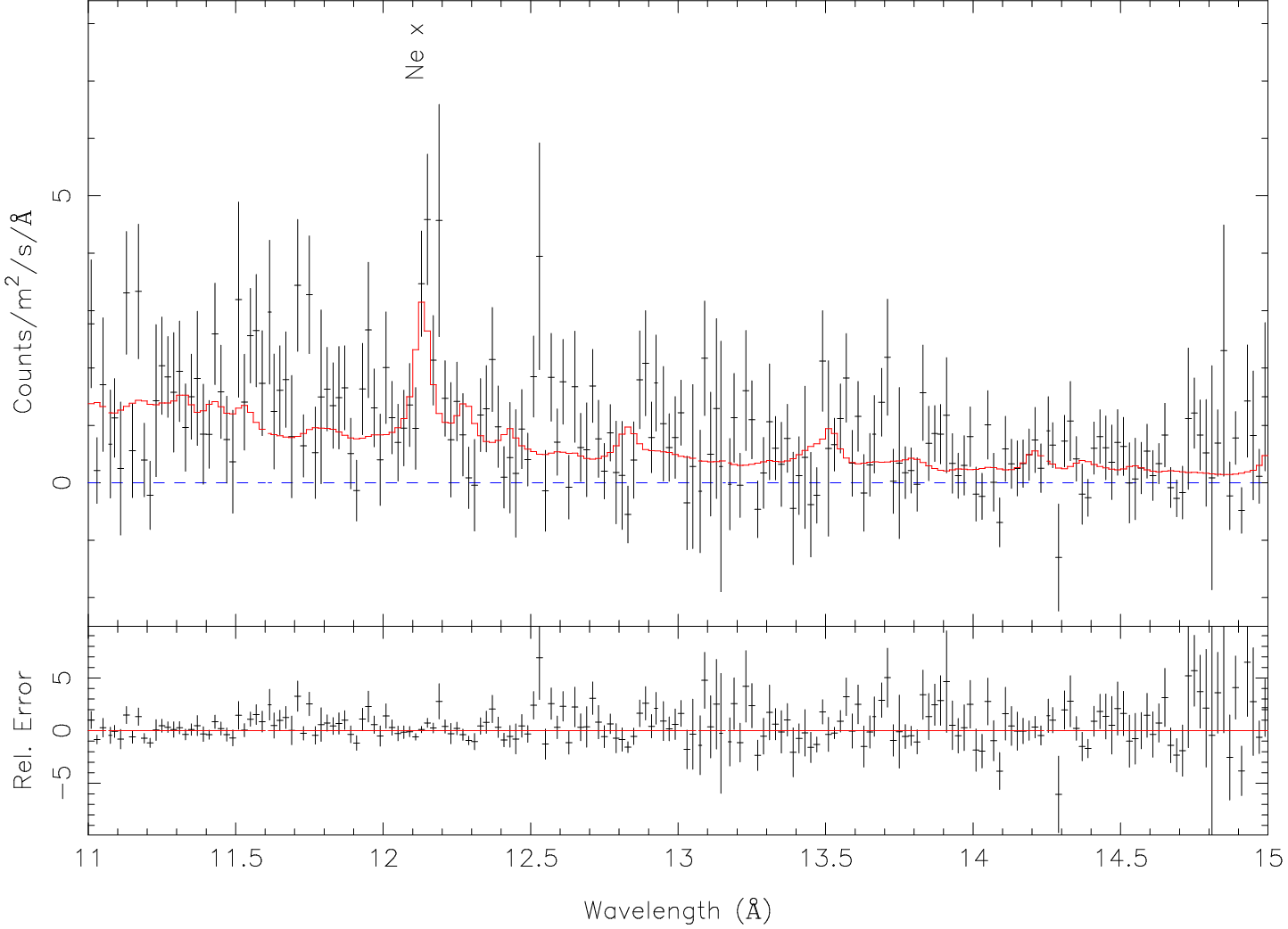}
  \caption{\emph{RGS} data, model and relative residuals for the orbital phase-resolved spectra.
\emph{From top to bottom}: eclipse, eclipse-egress and egress in the 0111010101 ID observation.
  }
  \label{fig:eclorbit}
\end{figure*}

To analyse spectral variations, the entire light curves were divided into different periods from which corresponding spectra were extracted (see figure~\ref{fig:lcurve}): eclipse period, eclipse-egress and egress for the first observation (ObsID 0111010101) and high count rate and low count rate for the second one (ObsID 0400550201). The source spectrum for each one was extracted from \emph{RGS} as described in Sect.~\ref{data}. The same model employed for analysing the average spectra was used in order to study the parameter variations (see Tables~\ref{tab:ecl-2cie} and \ref{tab:noecl}). In figures~\ref{fig:eclorbit} and \ref{fig:hcr}, \emph{RGS} orbital phase-resolved spectra are shown.

\subsubsection{ObsID 0111010101}
\label{orbphase1}

\begin{figure}[htb]
  \centering
  \includegraphics[angle=-90,width=\columnwidth]{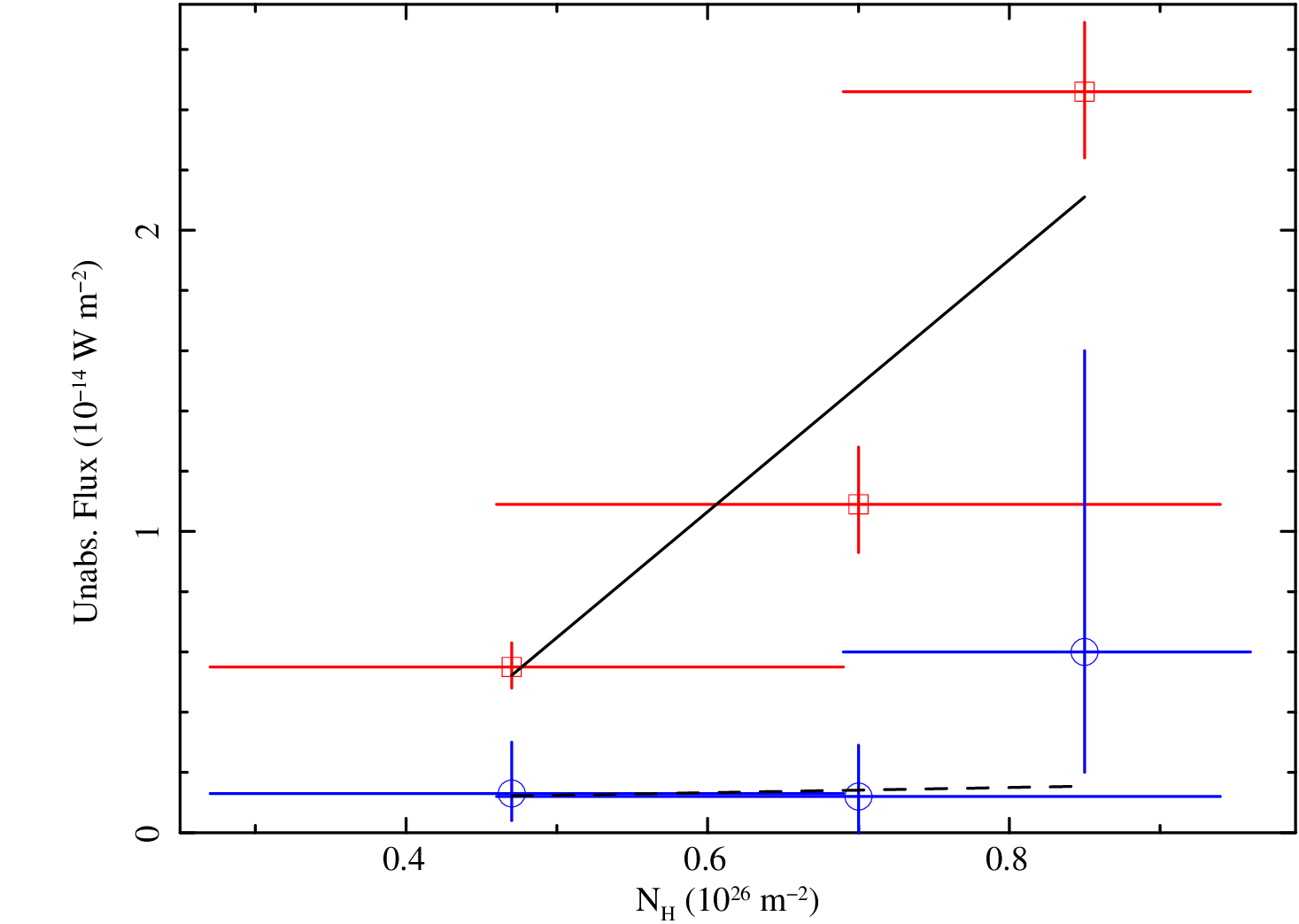}
  \caption{Correlation of the unabsorbed flux (red square corresponds to \emph{Cie}1 and blue circle corresponds to \emph{Cie}2) versus the column density $N_H$.
  }
  \label{fig:NHvsFlux}
\end{figure}

\begin{figure*}[h!tb]
  \centering
  \includegraphics[angle=-90,width=\columnwidth]{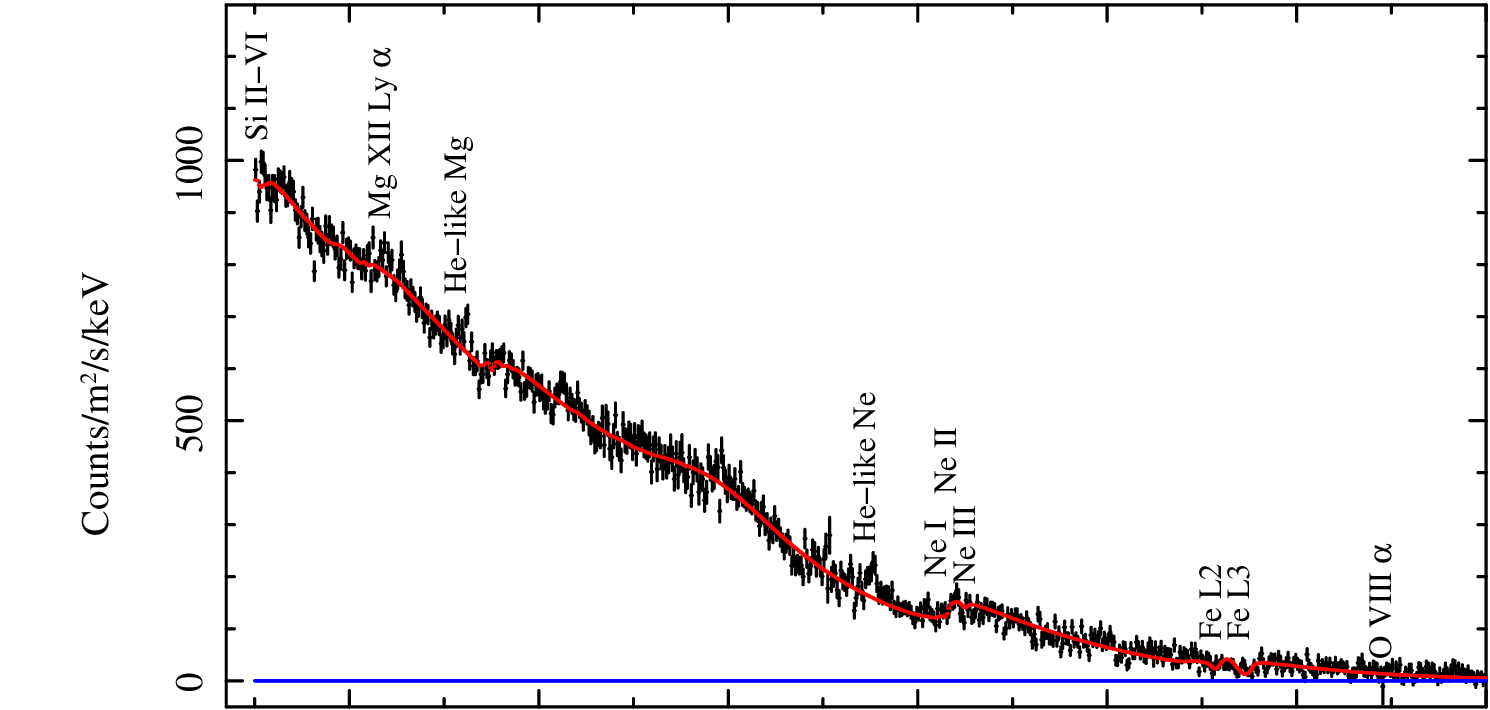}
 \includegraphics[angle=-90,width=\columnwidth]{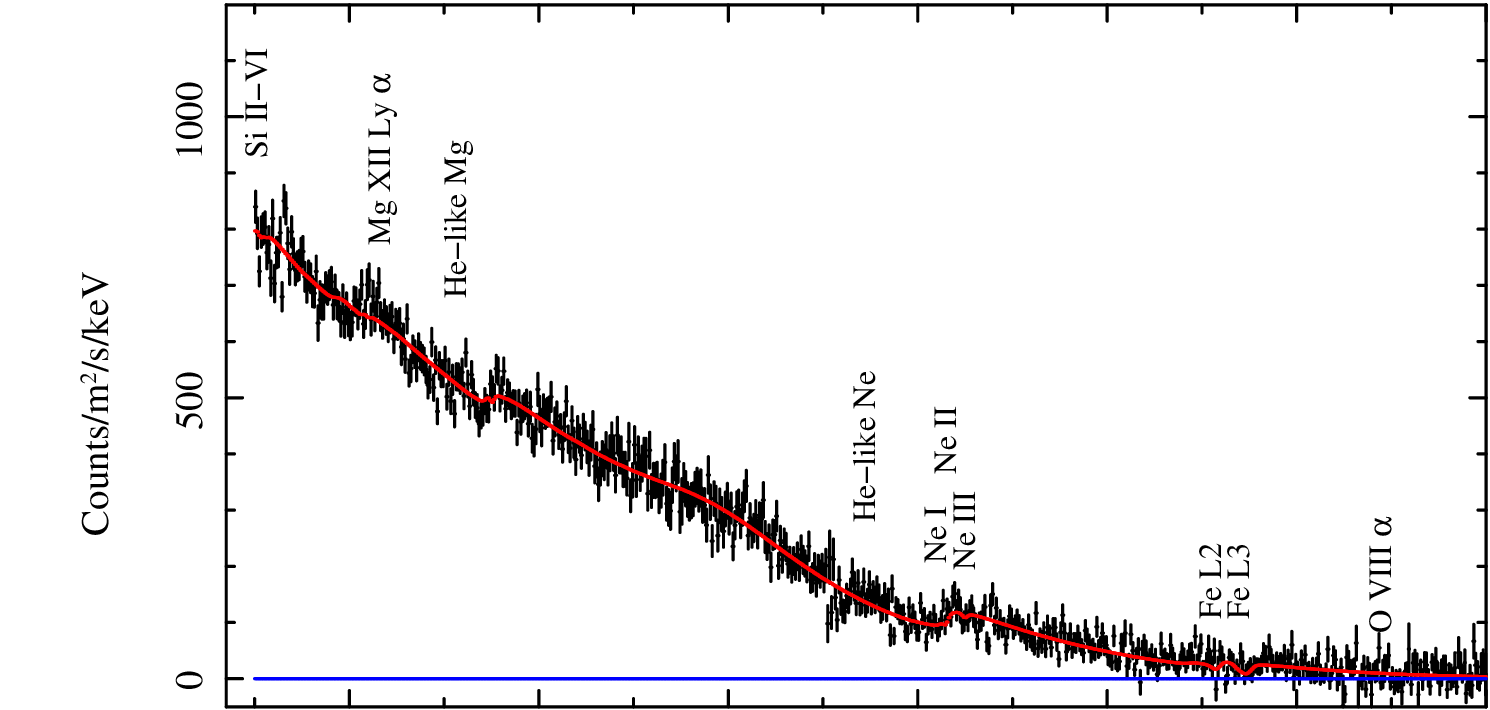}
 \includegraphics[angle=-90,width=\columnwidth]{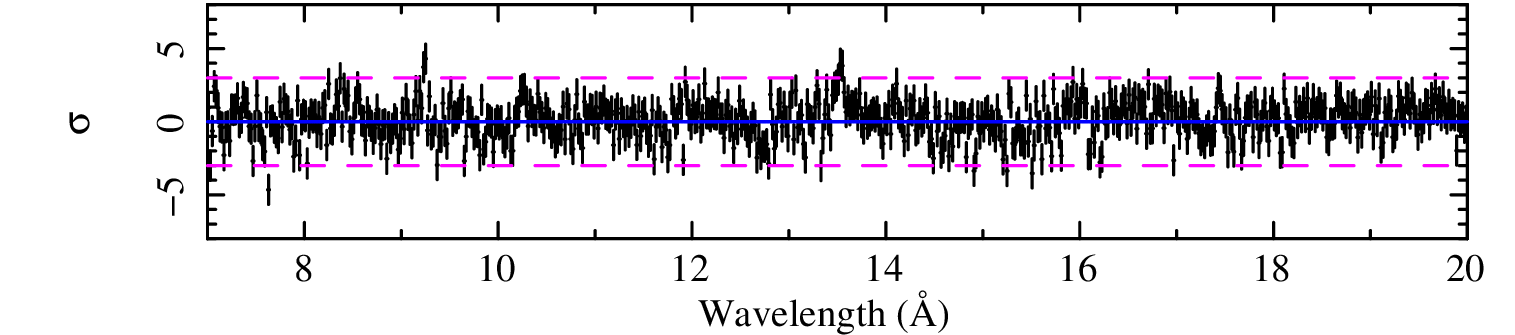}
 \includegraphics[angle=-90,width=\columnwidth]{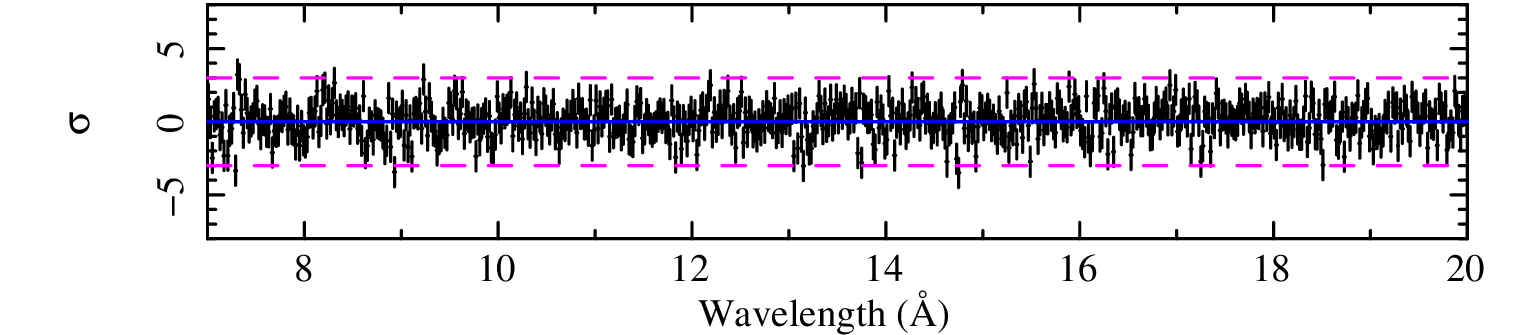}

  \caption{\emph{RGS} data, model and residuals in terms of standard deviations for the high (left panels) and low (right panels) count rate spectra in the 0400550201 ID observation.
  }
  \label{fig:hcr}
\end{figure*}

Due to the limited counts in the spectrum during the eclipse orbital phase, the orbital phase-resolved spectrum exhibits substantial uncertainties in the model parameters, and as a consequence the majority of these parameters remain consistent between them within the error range. Nevertheless, a clear progressive increase both in the emission measure and flux of component \emph{cie}1 can be derived, meanwhile these parameters of component \emph{cie}2 seems to be constant from the beginning to the middle of the observation and then to increase to the egress phase. The evolution of the hydrogen column density of the absorbing material reflects that the amount of surrounding material increases in the line of sight towards the neutron star as it emerges from the eclipse. An enhancement is present from eclipse to egress. On the other hand, the $N_H$ remains larger than the Galactic absorption value throughout the observation. This indicates the presence of extra matter either stellar wind, disk around the compact object or both. An apparent linear correlation between column density and unabsorbed flux was found (see Fig.~\ref{fig:NHvsFlux}).

Hydrogenic-like lines of Mg \textsc{xii} and Ne \textsc{x} seems to be present along the eclipse-egress transition showing a little variability but no Helium-like lines were clearly detected in the orbital phase-resolved spectra suggesting that the region where are produced should be in the surroundings of the NS. The emission lines were described by Gaussian components and the resulting line parameters are listed in Table~\ref{tab:lines-evolve}. In the eclipse spectrum the EWs of these lines are $\sim$11.7 eV and $\sim$23.3 eV; in the eclipse-egress spectrum, $\sim$5.9 eV and $\sim$36.4 eV; and in the egress spectrum, $\sim$34.8 eV and $\sim$73.5 eV. It should be noted that large uncertainties for the emission line fluxes prevents us from analysing the EW evolution and drawing any strong conclusions. In Figure~\ref{fig:eclorbit} time-resolved spectra fits and residuals are shown.

\subsubsection{ObsID 0400550201}
\label{orbphase2}

The behaviour of the source during the out-of-eclipse orbital phase also showed emission changes between the low and high fluxes (relative to the mean). The flux differences between them were statistically significant at one sigma. Assuming that the power-law component $PL_1$ is related to the direct X-ray emission, and the increase from HCR to LCR of the column density [$N_H (\mathrm{LCR}) \sim 1.5\,N_H (\mathrm{HCR})$], the X-ray flux rise could be due to the fall in $N_H$ although it cannot be ruled out an increment in mass accretion rate too. The scatter emission due to the interaction with the stellar wind and/or the accretion disk should be described by the another power-law component $PL_2$ enhanced by the broad Gaussian component. The power-law ($PL_2$) unabsorbed flux is slightly larger in the HCR state than in the LCR state and the enlargement produced by the Gaussian component is equivalent to each other taking the uncertainties into account.

In the HCR spectrum, both the He-like Mg and Ne triplets are detected as a single feature centred at the intercombination wavelength. The Mg \textsc{xii} Ly $\alpha$ is blue-shifted and the Ne \textsc{x} Ly $\beta$ is red-shifted with respect to the reference wavelengths. The line parameters are listed in Table~\ref{tab:gaus}. On the contrary, none of these features are present in the LCR spectrum.

Nevertheless, residuals around the He-like Ne \textsc{ix} triplet in the HCR spectrum suggest that this emission line could be separated by its components. A local fit was adopted to obtain line parameters which are also listed in Table~\ref{tab:ne}. The forbidden to intercombination line ratio derived from the HCR \emph{RGS} spectrum for helium-like neon ion is $\mathcal{R}_\mathrm{obs} = f/i = 0.24^{+0.18}_{-0.24}$ and it is totally consistent with the HCR average spectrum value. The forbidden and intercombination to resonance line ratio derived is $\mathcal{G}_\mathrm{obs}=(f+i)/r=3.6$ and a relative error equivalent to 100\%. According to the theoretical wavelength from \citet{1987JPCRD..16S....K} and \citet{2001ApJS..134..331D} the individual velocity shifts derived from the averaged and HCR spectra for resonance line of 960 km/s and 670 km/s, for intercombination line of $-$200 km/s and $-$380 km/s, and for forbidden line of $-$660 km/s and $-$220 km/s, respectively.

\begin{figure}[h!tb]
  \centering
  \includegraphics[angle=-90,width=\columnwidth]{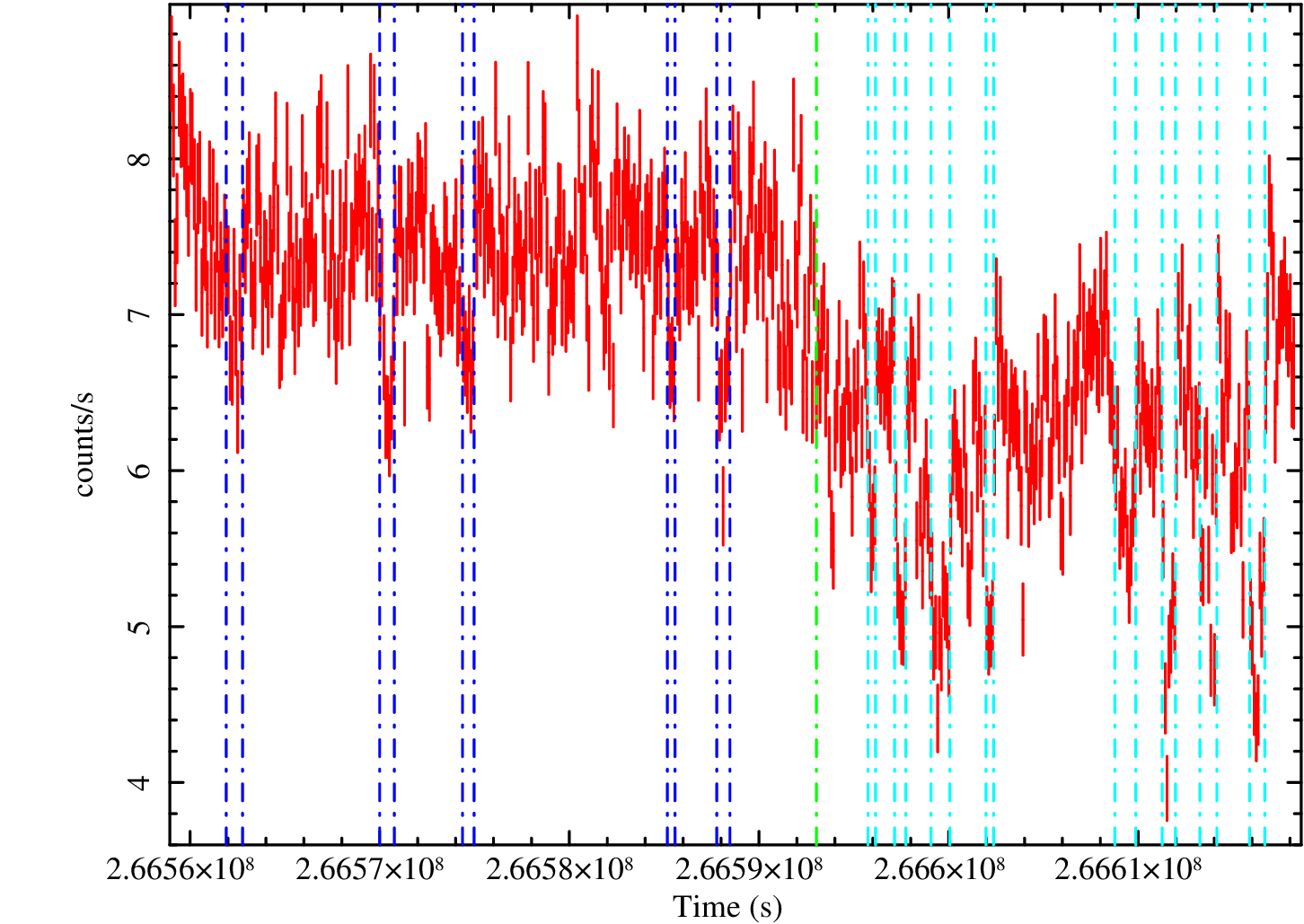}
  \caption{The dot-dash green vertical line indicates the HCR (on the left) and LCR (on the right) time intervals. The dark blue and light blue dot-dash vertical lines indicate the dips from the HCR and LCR time intervals, respectively.
  }
  \label{fig:dips-nodips}
\end{figure}

On the other hand, the evolution of the light curve in the out-of-eclipse observation showed sudden dips in both time-resolved spectra, as can be seen in Fig.~\ref{fig:lcurve} (right panel). Therefore, good time intervals for defining the dip and non-dip intervals were derived and spectra were extracted from these defined intervals. The duration of the dip events are in the range (200, 1300) s with an accumulated time of $\sim$7.9 ks for the HCR spectrum and (400, 1100) s with an accumulated time of $\sim$4.9 ks for the LCR spectrum, as it is indicated in Fig.~\ref{fig:dips-nodips}. The same component models were applied in order to compare the model parameters with the previous fits (see Table~\ref{tab:noecl}). Significant changes in the model parameters were found showing a complex interaction between X-ray emitter, stellar wind and accretion disk or other material distribution around the neutron star. Dips could be explained by an overall rise in absorption but here the hydrogen column density is found to be $N_H = 3.5^{+0.5}_{-0.4}$ in the high flux state (smaller than that in the HCR and out-of-dips averaged spectra) and $N_H = 7.8^{+1.4}_{-1.1}$ in the low flux state (similar to the value obtained in the LCR and out-of-dips averaged spectra). Therefore, these results suggest that dips can be attributed to a drop in the mass accretion rate. It should be noted that the X-ray emission in the LCR-dip events did not need the broad Gaussian component to describe this spectrum. Both dip durations and hardness light curves from \emph{PN} instrument showed a similar behaviour as \emph{RGS} data \citep[e.g. see Fig.~3 and 6 in][and Fig.~\ref{fig:lcurve} in this work]{2021MNRAS.501.5892S}. They concluded that the observed dips were probably due to inevitable instabilities occurring at the inner edge of the disk interacting with the NS magnetosphere.

\section{Discussion}
\label{discussion}

The eclipse spectrum of Cen X$-$3 has usually been described by absorbed power-laws and a variable number of emission (Gaussian) lines in the \emph{RGS} wavelength range \citep{1996PASJ...48..425E,2003ApJ...582..959W}, 
although it was applied to fit a broader wavelength range (1.2-17.7) \AA\, compared to the \emph{RGS} wavelength band (7.0-15.0) \AA. Generally, it is assumed that X-ray source photoionises the strong stellar wind of the giant/supergiant companion where these emission lines are produced. \citet{2003ApJ...582..959W} fitted the eclipse spectrum of Cen X-3 with pure collisionally ionised plasma in the energy range [0.7--10.0] keV. Although this model reproduced most of the observed emission lines and also accounted for all of the continuum emission below $\sim 4$ keV, they rejected this model and adopted a pure photoionised plasma. However, they did not attempt a hybrid scenario of collisional and photoionised plasma. SPEX photoionised plasma and photoionised absorbed models as well as a hybrid collisional and photoionised plasma were used to describe the spectra, but the fits did not describe the observed emission lines well. The most prominent emission lines were the Ly $\alpha$ transitions from the hydrogen-like ions Mg and Ne which can be produced in a collisional plasma. A combination of two absorbed collisional ionisation plasma models reproduced most of the observed emission lines and also accounted for all of the continuum emission. The level of counts is not large enough to make useful constraints on the components of the helium-like triplets of Mg \textsc{xi} or Ne \textsc{ix}, but an apparent enhancement of the forbidden line could be present. Furthermore, the hydrogen-like emission lines of Mg \textsc{xii} and Ne \textsc{x} were clearly detected with a significance $\sim 5\, \sigma$ in our eclipse spectrum. Similar behaviour of these lines were also reported by \citep{1996PASJ...48..425E}, using \emph{ASCA} data, and \citep{2003ApJ...582..959W}, using \emph{Chandra/MEG/HEG} data. No other emission lines were reported by these authors in the wavelength band we analysed from \emph{RGS} where Ne \textsc{x} at 10.240 \AA\, and Fe \textsc{xviii} at 10.627 \AA\, could be present. Cen X$-$3 was also observed with the \emph{Chandra} satellite on 2000 December 30 using \emph{HETGS} \citep{2005ApJ...634L.161I} (just one month before the first \emph{XMM-Newton} observation and covering the same orbital phase interval, excluding the eclipse egress). Remarkably, these authors did not report any detection of hydrogen-like emission lines, probably because they focused their analysis in the iron complex and the Si \textsc{xiv} emission line. Orbital phase-resolved spectra were also fitted with the same model we defined on the average spectrum. The H-like emission lines showed changes throughout the eclipse observation, although no other emission lines could be clearly distinguished from the continuum emission.

On the other hand, during the out-of-eclipse orbital phases, the neutron star is in direct line of sight with the observer and the averaged spectrum shows a high X-ray continuum emission. On average, the source flux was around 50 times greater than in eclipse from 7 \AA\, to 11 \AA, near 30 times from 11 \AA\, to 15 \AA\, and falls to the X-ray background flux above 20 \AA. The continuum model was a highly absorbed power law component to describe the X-rays from the neutron star and a less-absorbed power law plus a broad Gaussian components to account the scattered emission in the extended wind. Emission lines with a significance of $\gtrsim 3\, \sigma$ were fitted with Gaussian line profiles. Residuals in the fit at 13.5 \AA\, suggested the presence of the He-like ion Ne \textsc{ix} and it was found that $\mathcal{R}_\mathrm{obs} = f/i = 0.3^{+0.4}_{-0.3}$ and $\mathcal{G}_\mathrm{obs}=(f+i)/r=2.5\pm2.4$. At the temperature corresponding to the maximum of the He-like Ne abundance for a collisional plasma, i.e. 2.0 MK, the $\mathcal{R}$ parameter points to electron density $\sim (2-7)\times 10^{12}$ cm$^{-3}$ but the curves for pure photoionised plasma and collisional plasma are indistinguishable \citep{2000A&AS..143..495P}. Another indication of a hybrid plasma might be the value of $\mathcal{G}_\mathrm{obs}$ which is consistent with a plasma temperature of 2.0 MK despite the large uncertainties \citep{2000A&AS..143..495P}. In a recent research, \citet{2024A&A...690A.360S} studied a ToO Chandra observation and reported the detection of He-like triplets Mg \textsc{xi} and Si \textsc{xiii} in the \emph{Chandra} spectra of Cen X$-$3. From the Mg \textsc{xi} and Si \textsc{xiii} ions $\mathcal{G}$ ratio values in Segment 3, they estimated a temperature of $\sim 7\times 10^5$ K and $\sim 7\times 10^6$ K, respectively \citep[see Fig.~12 and Table~6 in][]{2024A&A...690A.360S}. From the He-like iron, \citet{2005ApJ...634L.161I} derived a temperature of the emitting region in the range $(1-4)\times 10^7$ K assuming a photoionised plasma. Unfortunately, this emission line was not resolved in the \emph{XMM-Newton} observation, therefore, it could not obtain an estimation of the temperature. Concerning the $\mathcal{R}$ ratio in Segment 3, the value of Mg \textsc{xi} implied an electron density of $\sim (1.5-5.0)\times 10^{12}$ cm$^{-3}$ but it could not be calculated for Si \textsc{xiii}. These values are of the same order of magnitude as the fit results found in this work. However, the electron densities they obtained in Segments 1 and 2 were one or two orders of magnitude greater than the previous one, indicating significant plasma density variations depending on the location and orbital phase. \citet{1996PASJ...48..425E} studied an observation with \emph{ASCA} (orbital phase range [-0.31--0.20]) obtaining an electron density $\sim 1.3\times 10^{11}$ cm$^{-3}$, in agreement with \citet[][\emph{Chandra} observation in the orbital phase range (0.13--0.40)]{2005ApJ...634L.161I}. Possible structures around the neutron star could be an accretion disc or an accretion wake with an electron density as high as $\sim 10^{12}$ cm$^{-3}$. It is also noticed that the equivalent hydrogen column derived in this work ($N_H \sim 10^{22}$ cm$^{-2}$) was one or two orders of magnitude smaller than the column density obtained by \citet{2024A&A...690A.360S}. Therefore, it seems that the accretion and reprocessing environment in Cen X-3 during the \emph{Chandra} observation was quite different when compared to \emph{XMM-Newton/RGS} results.

Ne \textsc{ix} triplet cannot be clearly distinguished from the X-ray background as well as Mg \textsc{xi} triplet in the eclipse spectrum. H-like ions of Mg \textsc{xii} and Ne \textsc{x} Ly $\beta$ and He-like ions of Mg \textsc{xi} and Ne \textsc{ix} were detected in the out-of-eclipse spectra with a significance of $\sim 3\, \sigma$ but it could not be well-resolved. The broad Gaussian component used to describe the continuum enhancement centred at $\sim 12$ \AA\, is probably the reason that did not allow us to detect the Ne \textsc{x} Ly $\alpha$ during the out-of-eclipse transition. Some new absorption features, compared to the eclipse-averaged spectrum, could be identified as Ne and Fe edges which should be part of the ISM matter in the line of sight of Cen X$-$3. No significant absorption lines were detected in our spectra with a significance $> 3\, \sigma$.

According to the light curve of the out-of-eclipse observation, X-ray flux fluctuated in short time-scales presenting pronounced dips through this orbital phase interval. Orbital phase-resolved spectroscopy revealed that the decrease in the X-ray flux during the dip intervals was due to a drop in the accretion rate rather than an increase of the column density absorption. Nevertheless, the LCR spectrum presented a higher intrinsic absorption than the HCR spectrum and the reduction in the X-ray flux could be produced by the rise in the column density in the direct emission ($PL_1$ component) and by a reduction in the accretion rate in the scattered emission ($PL_2$ component). It is also noted that the radiation emitted by the broad Gaussian component disappeared totally in the LCR-dip spectrum, while remain nearly constant in the other spectra taking the uncertainties into account.

\section{Conclusion}
\label{conclusion}

We have analysed two \emph{XMM-Newton/RGS} both averaged and resolved spectra of Cen X$-$3 taken in two complementary orbital phase intervals, although separated about six years between them. From the light curves, we derived the orbital phase range of the observations and did time resolved spectroscopy to analyse the changes in the emission lines and in the continuum parameters with the orbital period. As expected, we needed different model components to fit each average spectrum and revealed significant different behaviour depending on the orbital phase of the Cen X$-$3 system. We found strong variations in source flux especially in the out-of-eclipse observation. The continuum emission during eclipse observation was reasonably fitted by two hot collisionally ionised plasmas which also explained the observed emission line of hydrogen-ions. During the out-of-eclipse observation, a phenomenological model was applied for fitting the X-ray emission. Any attempt to describe these spectra by a photoionised plasma failed or had a worse fit compared to the models used in this work. However,it is quite probably that a mixture of collisional and photoionised plasmas may be present in this X-ray binary system. To confirm this hypothesis it should be observed a high-resolution and high signal-to-noise X-ray spectrum to detect as many He-like emission lines as possible and resolve the resonant, intercombination and forbidden components.

\emph{RGS} could clearly detect emission lines from the hydrogenic ions of Ne and Mg and showed evidence of other emission lines such as Si \textsc{v}, Mg \textsc{xi}, Ne \textsc{x} (10.240 \AA) and Fe \textsc{xviii} (10.627 \AA), although difficult to distinguish from the X-ray continuum in the eclipse-egress spectrum. Although the helium-like triplet of neon seemed to be present in both spectra it could not be clearly distinguished from the X-ray background in the eclipse spectrum. On the other hand, the line ratios, $\mathcal{R}_\mathrm{obs}$ and $\mathcal{G}_\mathrm{obs}$, derived from the out-of-eclipse were consistent with the ratios obtained by \citet{2005ApJ...634L.161I} and \citet{2024A&A...690A.360S}. The He-like ions such as Mg \textsc{xi} or Ne \textsc{ix} provide useful diagnostics for X-ray emitting plasma. Unfortunately, \emph{RGS} could not resolve/detect these triplets in both observations and, therefore, we could not estimate the average distances from the stellar surface of these ions. Simultaneous observations with \emph{RGS} and \emph{HETG} should be interesting to investigate He-like triplet lines for improving the knowledge of the accretion/reprocessing plasma in the surrounding of the neutron star. Moreover, Cen X$-$3 is among the main sources to be observed by \emph{XRISM} opening a new era in high-resolution X-ray spectroscopy thanks to the microcalorimeter instrumentation.

Other absorption features detected in the out-of-eclipse spectrum suggested that should be formed in the ISM in the line-of-sight in the direction of Cen X$-$3. However, near the Fe L-edge the \emph{RGS} absorption lines blend and may be similar and degenerate by features produced by other elements or ions \citep{2013A&A...551A..25P}.

\begin{acknowledgements}

  The author(s) acknowledge the financial support from the MCIN with funding from the European Union NextGenerationEU and Generalitat Valenciana in the call Programa de Planes Complementarios de I+D+i (PRTR 2022). Project (Athena-XIFU-UA), reference ASFAE/2022/002. J.J.R.R. acknowledges financial support from the Spanish Ministry of Education, Culture and Sport fellowship PRX23/00270, and also thanks all the staff from SRON for their collaboration and hospitality there. We would like to thank particularly Liyi Gu and Jelle de Plaa for invaluable assistance in analysing \emph{RGS} data. The research leading to these results has received funding from the European Union's Horizon 2020 Programme under the AHEAD project (grant agreement n. 654215). We acknowledge the constructive report of the referee whose comments helped us to improved the content of this paper. Part of this work has been developed in the framework of the PID2024-155779OB-C33 project.
\end{acknowledgements}

\bibliographystyle{aa}
\bibliography{sron2017_rgs01}

\newpage

\onecolumn

\begin{appendix}

\section{Emission lines}

\begin{table*}[h!tb]
\caption{Parameters of the emission lines according to the fit results from the three models.
}
\label{tab:lines-evolve}
\centering
\begin{adjustbox}{max width=0.95\textwidth}\begin{tabular}{lccccccccc}
\midrule
\midrule
 & \multicolumn{3}{c}{Eclipse} & \multicolumn{3}{c}{Eclipse-egress} & \multicolumn{3}{c}{Egress} \\
 & \multicolumn{3}{c}{Model component} & \multicolumn{3}{c}{Model component} & \multicolumn{3}{c}{Model component} \\
Parameter & \emph{pion} & \emph{cie} & \emph{Gaussian} & \emph{pion} & \emph{cie} & \emph{Gaussian} & \emph{pion} & \emph{cie} & \emph{Gaussian} \\
\midrule
 & \multicolumn{9}{c}{Mg \textsc{XII} Ly $\alpha$} \\ \midrule
\multirow{2}{*}{Flux (10$^{41}$ ph\,s$^{-1}$)} & $3.2$ & $2.8$ & \multirow{2}{*}{$9^{+6}_{-3}$} & $3.9$ & $3.2$ & \multirow{2}{*}{$8^{+10}_{-5}$} & $13.3$ & $7.6$ & \multirow{2}{*}{$20^{+40}_{-10}$} \\
 & $1.6$ & $1.6$ &  & $1.9$ & $1.8$ & & $6.6$ & $4.2$ & \\
\multirow{2}{*}{$\lambda$ (\AA)} & $8.41921$ & $8.419$ & \multirow{2}{*}{$8.443^{+0.016}_{-0.014}$} & $8.41921$ & $8.419$ & \multirow{2}{*}{$8.424^{+0.020}_{-0.021}$} & $8.41921$ & $8.419$ & \multirow{2}{*}{$8.427^{+0.006}_{-0.008}$} \\
 & $8.42462$ & $8.424$ &  & $8.42462$ & $8.424$ &  & $8.42462$ & $8.424$ & \\
\multirow{2}{*}{FWHM (\AA)} & $0.007$ & $0.005$ & \multirow{2}{*}{$0.005$ (fixed)} & $0.007$ & $0.006$ & \multirow{2}{*}{$0.005$} & $0.007$ & $0.005$ & \multirow{2}{*}{$0.005$} \\
 & $0.007$ & $0.005$ &  & $0.007$ & $0.006$ &  & $0.007$ & $0.005$ & \\
\midrule
 & \multicolumn{9}{c}{Mg \textsc{XI} (r i f)} \\ \midrule
\multirow{3}{*}{Flux (10$^{41}$ ph\,s$^{-1}$)} & $2.2$ & $0.4$ & \multirow{3}{*}{$8^{+9}_{-4}$} & $2.8$ & $0.4$ & \multirow{3}{*}{$17^{+15}_{-9}$} & $5.9$ & $2.5$ & \multirow{3}{*}{$12^{+5}_{-3}$} \\
 & $0.7$ & $0.07$ &  & $0.4$ & $0.07$ &  & $0.5$ & $0.4$ &  \\
 & $2.8$ & $0.3$ &  & $2.5$ & $0.3$ &  & $2.1$ & $1.5$ & \\
\multirow{3}{*}{$\lambda$ (\AA)} & $9.16875$ & $9.170$ & \multirow{3}{*}{$9.21^{+0.03}_{-0.02}$} & $9.16875$ & $9.170$ & \multirow{3}{*}{$9.289^{+0.023}_{-0.018}$} & $9.16875$ & $9.170$ & \multirow{3}{*}{$9.206^{+0.011}_{-0.009}$} \\
 & $9.23121$ & $9.232$ &  & $9.23121$ & $9.232$ &  & $9.23121$ & $9.232$ &  \\
 & $9.31434$ & $9.315$ &  & $9.31434$ & $9.315$ &  & $9.31434$ & $9.315$ & \\
\multirow{3}{*}{FWHM (\AA)} & $0.007$ & $0.006$ & \multirow{3}{*}{$0.005$ (fixed)} & $0.007$ & $0.006$ & \multirow{3}{*}{$0.005$} & $0.008$ & $0.005$ & \multirow{3}{*}{$0.005$} \\
 & $0.007$ & $0.006$ & & $0.007$ & $0.006$ &  & $0.008$ & $0.005$ &  \\
 & $0.007$ & $0.006$ & & $0.007$ & $0.006$ &  & $0.008$ & $0.005$ &  \\
\midrule
 & \multicolumn{9}{c}{Ne \textsc{X} Ly $\beta$} \\ \midrule
\multirow{2}{*}{Flux (10$^{41}$ ph\,s$^{-1}$)} & $0.2$ & \multirow{2}{*}{$1.7$} & \multirow{2}{*}{------} & $0.9$ & \multirow{2}{*}{$2.0$} & \multirow{2}{*}{------} & $2.7$ & \multirow{2}{*}{$4.8$} & \multirow{2}{*}{------} \\
 & $0.2$ & & & $0.8$ & & & $2.0$ &  &  \\
\multirow{2}{*}{$\lambda$ (\AA)} & $10.2390$ & \multirow{2}{*}{$10.240$} & \multirow{2}{*}{------} & $10.2390$ & \multirow{2}{*}{$10.240$} & \multirow{2}{*}{------} & $10.2390$ & \multirow{2}{*}{$10.240$} & \multirow{2}{*}{------} \\
 & $10.2396$ &  &  & $10.2396$ &  &  & $10.2396$ &  &  \\
\multirow{2}{*}{FWHM (\AA)} & $0.008$ & \multirow{2}{*}{$0.007$} & \multirow{2}{*}{------} & $0.008$ & \multirow{2}{*}{$0.008$} & \multirow{2}{*}{------} & $0.009$ & \multirow{2}{*}{$0.005$} & \multirow{2}{*}{------} \\
 & $0.008$ &  &  & $0.008$ & &  & $0.009$ &  &  \\
\midrule
 & \multicolumn{9}{c}{Ne \textsc{X} Ly $\alpha$} \\ \midrule
\multirow{2}{*}{Flux (10$^{42}$ ph\,s$^{-1}$)} & $1.3$ & $1.0$ & \multirow{2}{*}{$30^{+40}_{-20}$} & $1.7$ & $1.2$ & \multirow{2}{*}{$50^{+200}_{-40}$} & $5.7$ & $2.9$ & \multirow{2}{*}{$8^{+50}_{-5}$} \\
 & $0.7$ & $0.6$ &  & $0.8$ & $0.7$ &  & $2.7$ & $1.6$ &  \\
\multirow{2}{*}{$\lambda$ (\AA)} & $12.1328$ & $12.1320$ & \multirow{2}{*}{$12.147^{+0.008}_{-0.010}$} & $12.1328$ & $12.1320$ & \multirow{2}{*}{$12.142^{+0.011}_{-0.008}$} & $12.1328$ & $12.1320$ & \multirow{2}{*}{$12.148^{+0.008}_{-0.005}$} \\
 & $12.1382$ & $12.1375$ &  & $12.1382$ & $12.1375$ & & $12.1382$ & $12.1375$ & \\
\multirow{2}{*}{FWHM (\AA)} & $0.010$ & $0.009$ & \multirow{2}{*}{$0.005$ (fixed)}& $0.010$ & $0.009$ & \multirow{2}{*}{$0.005$} & $0.010$ & $0.007$ & \multirow{2}{*}{$0.005$} \\
 & $0.010$ & $0.009$ &  & $0.010$ & $0.009$ &  & $0.010$ & $0.007$ &  \\
\midrule
 & \multicolumn{9}{c}{Ne \textsc{IX} (r i f)} \\ \midrule
\multirow{3}{*}{Flux (10$^{41}$ ph\,s$^{-1}$)} & $2.1$ & $0.5$ & \multirow{3}{*}{------} & $2.3$ & $0.5$ & \multirow{3}{*}{------} & $8.7$ & $2.6$ & \multirow{3}{*}{------} \\
 & $0.7$ & $0.08$ & & $0.4$ & $0.08$ & & $0.5$ & $0.4$ &  \\
 & $2.6$ & $0.3$ & & $1.7$ & $0.3$ & & $2.0$ & $1.8$ &  \\
\multirow{3}{*}{$\lambda$ (\AA)} & $13.4471$ & $13.448$ & \multirow{3}{*}{------} & $13.4471$ & $13.448$ & \multirow{3}{*}{------} & $13.4471$ & $13.448$ & \multirow{3}{*}{------} \\
 & $13.5529$ & $13.553$ & & $13.5529$ & $13.553$ & & $13.5529$ & $13.553$ &  \\
 & $13.6987$ & $13.700$ & & $13.6987$ & $13.700$ & & $13.6987$ & $13.700$ &  \\
\multirow{3}{*}{FWHM (\AA)} & $0.011$ & $0.010$ & \multirow{3}{*}{------} & $0.011$ & $0.010$ & \multirow{3}{*}{------} & $0.011$ & $0.008$ & \multirow{3}{*}{------} \\
 & $0.011$ & $0.010$ & & $0.011$ & $0.010$ & & $0.011$ & $0.008$ &  \\
 & $0.011$ & $0.010$ & & $0.011$ & $0.010$ & & $0.011$ & $0.008$ &  \\
\midrule
\midrule
\end{tabular}
\end{adjustbox}
\end{table*}

\end{appendix}

\end{document}